\documentclass[showpacs,preprintnumbers,prd]{revtex4}

\input{epsf}

\newcommand{\be}{\begin{equation}}
\newcommand{\ee}{\end{equation}}
\newcommand{\bea}{\begin{eqnarray}}
\newcommand{\eea}{\end{eqnarray}}
\newcommand{\bef}{\begin{figure}}
\newcommand{\eef}{\end{figure}}

\def\H#1{{\rm ^{#1}H}}
\def\He#1{{\rm ^{#1}He}}
\def\Li#1{{\rm ^{#1}Li}}
\def\Be#1{{\rm ^{#1}Be}}
\def\G{\gamma}
\def\h#1{$^{#1}$H}
\def\he#1{$^{#1}$He}
\def\li#1{$^{#1}$Li}
\def\be#1{$^{#1}$Be}

\def\Obh2{$\Omega_b h^2$}

\newcommand{\simge}{\gtrsim}
\newcommand{\simle}{\lesssim}

\def\apj{Astrophys. J.}

\def\eps@scaling{0.70}

\def\showone#1{
  \centering
  \leavevmode
  \epsfxsize=\eps@scaling\linewidth
  \epsfbox{#1.eps}
}

\def\epstwo@scaling{0.48}

\def\showtwo#1#2{
  \centering
  \leavevmode
  \epsfxsize=\epstwo@scaling\linewidth
  \epsfbox{#1.eps} \hfil
  \epsfxsize=\epstwo@scaling\linewidth
  \epsfbox{#2.eps}
}

\begin{document}

\title{Big Bang Nucleosynthesis Constraints 
on Hadronically and Electromagnetically
Decaying Relic Neutral Particles} 
\author{Karsten Jedamzik} 
\affiliation{Laboratoire de Physique Math\'emathique et
Th\'eorique, Universit\'e de Montpellier II, 34095 Montpellier Cedex
5, France}

\begin{abstract}
Big Bang nucleosynthesis in the presence of 
decaying relic neutral particles
is examined in detail. 
All non-thermal processes 
important for the determination of light-element
abundance yields of \h2, \h3, \he3, \he4, \li6, and \li7 are coupled to 
the thermonuclear fusion reactions to obtain comparatively accurate results.
Predicted light-element yields are compared to observationally
inferred limits on primordial light-element abundances 
to infer constraints
on the abundances and properties of relic decaying particles with decay
times in the interval $0.01\, {\rm sec}\simle \tau_X\simle 10^{12}{\rm sec}$.
Decaying particles are typically constrained at early times by \he4 or \h2, 
at intermediate times by \li6, and at large times by the \he3/\h2 ratio. 
Constraints are shown for a large number of hadronic branching ratios and
decaying particle masses and may be applied to constrain the evolution of
the early Universe. 
\end{abstract}


\maketitle

\setlength{\baselineskip}{12pt}

\section{Introduction}

The epoch of Big Bang nucleosynthesis (BBN) is one of the furthest
reaching back probe of early cosmological conditions. It has thus
been invaluable, for example, for the realization that baryonic matter 
contributes only a small fraction to the present critical density, 
thus providing independent evidence for the existence of dark matter, or 
that extensions of the standard model of particle physics which include
further light degrees of freedom are observationally disfavored.
Due to its usefulness
BBN has thus been analysed in many variants (inhomogeneous BBN, BBN with
leptonic chemical potentials, BBN with antimatter domains, BBN with
decaying particles or evaporating primordial black holes, BBN with
Brans-Dicke gravity, etc. - for
reviews the reader is referred to Ref.~\cite{NBBNreview}). These studies have
been important for the
realization that the Universe must have been fairly close during the 
BBN epoch ($\sim 1$-$10^4$sec) to that described by the standard model
of BBN (SBBN). An SBBN scenario assumes a homogeneous baryonic plasma
free of defects or decaying relics and with close-to-vanishing leptonic
chemical potentials at constant baryon-to-entropy. With the recent accurate
determination of the fractional contribution of baryons to the critical
density $\Omega_bh^2 = 0.0223^{+0.0007}_{-0.0009}$ 
($h$ is the Hubble constant in units $100\,$ km s$^{-1}$Mpc$^{-1}$)
by the WMAP satellite mission~\cite{WMAP}
predicted primordial
light-elements in a SBBN scenario may be compared with those
observationally inferred. This comparison is good when \h2/H is considered,
fair when \he4, and inconclusive when \he3 is considered, 
but factor $2-3$ discrepant in the
case of \li7. It is not clear if this latter discrepancy is due to
stellar \li7 destruction effects, either in 
Pop II~\cite{Pinn,Richards,Lidepletion,Thea:01}
stars where it is
observed, or in a prior Pop III generation~\cite{Piau}, or indeed due to
a required modification of BBN, such as a population of decaying 
relics~\cite{J04-01,Jetal05}.
This discrepancy, however, is not subject of the present paper.

The current study presents in detail analysis and results of BBN with 
decaying neutral particles. Such scenarios may easily emerge, for example, in
supersymmetric extensions of the standard model, where relic
gravitinos produced at high temperatures decay during BBN~\cite{gravitino}. 
Similarly,
when the gravitino is the lightest supersymmetric particle, next-to-lightest
supersymmetric binos or sneutrinos may decay into gravitinos during or after BBN.
Many such BBN studies have been presented before, from
the pioneering works in the eighties~\cite{Balestra,Lindley,decay80s,Dimo,RS} 
to improved and updated analysis 
in the nineties~\cite{decay90s,J00,moredecay90s} 
up to very recent publications~\cite{Cyburt,J04-01,KKMlett,KKM}. 
Nevertheless, except for
Ref.~\cite{J04-01,KKMlett,KKM}
none of these works were able to properly treat the hadronic decay
of a relic particle {\it during} BBN. 

The purpose of this paper is twofold. It is intended to present a
catalog of constraints
on hadronically and electromagnetically decaying neutral particles over a wide range
of decay times ($10^{-2}$-$10^{12}$sec) as well as hadronic
branching ratios and relic particle masses. Prior studies have
presented results only for a very small number of hadronic branching ratios.
It should be noted that due to nonlinearities, either at smaller decay
times ($\tau\simle 10^4$sec) when thermal nuclear reactions still proceed
or due to interference of electromagnetic destruction and hadronic
production, for example, a simple interpolation between results at different
hadronic branching ratios often leads to erroneous or inaccurate results.
The second purpose of this study is to present the detailed analysis 
underlying the results presented in 
Ref.~\cite{J04-01,Jedamzik:2004ip,Cerdeno:2005eu,Jetal05}.
In general the philosophy in the analysis is to stay as close as possible
to experimental data. Furthermore, a large number (93, cf. 
Table~\ref{bigtable}) 
of processes are
included, all those which the author believes to play a role at the more
than a few per cent level. Particular attention is given to an as accurate
as possible determination of the synthesized \li6 in such scenarios. This
is also of importance as \li6 has been recently observed in surprisingly
large abundance
in about 10 Pop II stars~\cite{Asplund} and is often a byproduct of 
decaying~\cite{Dimo,J00} or annihilating~\cite{Jedamzik:2004ip} 
relic particles. Finally, it is noted that the present limits do not
apply to decaying electrically charged relic particles with
$\tau_{X}\simge 100\,$sec
as in this case, the existence of bound states between nuclei and the
relics may significantly change nuclear reaction rates~\cite{bound}.

The outline of the paper is as follows. In Sec. II a brief overview over the
physics of BBN in the presence of 
electromagnetically and hadronically decaying particles is presented.
Sec. III discusses observationally inferred primordial
light-element abundances and
derives the constraints on those as applied in the analysis. Sec. IV then
presents the constraints which apply to the abundance of putative relic
decaying particles, whereas conclusions are drawn in Sec. V. For the in
details interested reader and for the benefit of the author, a
number of appendices summarizes the analysis. This includes general
considerations (App. A), the numerical procedure (App. B), employed
kinematic relations (App. C), non-thermal electromagnetic interactions
(App. D), non-thermal hadronic interactions (App. E), and the thermal nuclear
reactions (App. F). 

\section{Overview of the Physics of BBN with Decaying Particles}
\label{sec:}

This section gives a brief and schematic account of the modifications to BBN 
when energetic particles are injected due to, for example, the decay of
relic particles. The decay channel of the relic determines the energy and
nature of the injected primaries. A BBN calculation with decaying particles
requires the detailed study of the thermalization of these primaries
in the primordial plasma, as well as that of secondaries 
produced during the thermalization process. Since for one energetic primary
several tens/hundreds 
of secondaries may be produced such calculations are sometimes
also referred to as cascade nucleosynthesis calculations.
In addition, thermal nuclear reactions operative
at early times ($\tau\simle 10^4$sec) have to be followed. One distinguishes
between primaries being either hadronic (nucleons, antinucleons, mesons), 
electromagnetic (photons, $e^{\pm}$), or inert (neutrinos or other weakly
interacting particles). This distinction is useful as the thermalization
of hadronically and electromagnetically interacting particles is very
different and thus impacts the light-element yields in different ways.
Both have nevertheless in common that over the whole range of decay
times considered here ($10^{-2}$-$10^{12}$sec) thermalization occurs
very rapidly on the Hubble scale. Redshifting of energetic particles has
therefore not to be considered. In contrast, inert particles, such as
neutrinos have typically no further interactions with the plasma and thus
also do not effect BBN yields. However, even neutrinos, 
if energetic enough ($\sim 100\,$GeV)
and injected very early on ($\tau\sim 1\,$sec) may not be regarded as inert
(cf. App. E.1), but since their effects are typically subdominant compared to 
hadronically and electromagnetically interacting particles, they are
not considered in the present study.

\bef
\showone{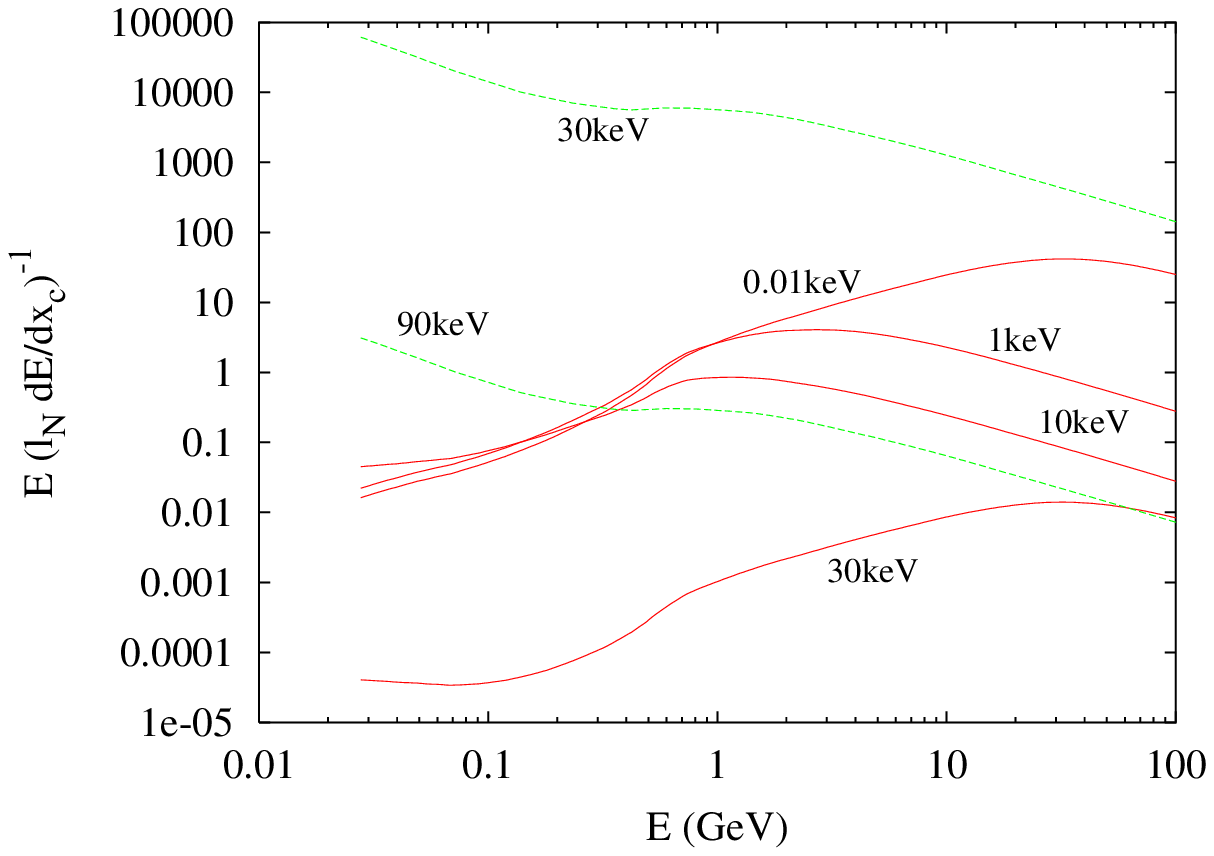}
\caption{The quantity $E (l_N |{\rm d}E/{\rm d}x|_c)^{-1}$ as a function of
nucleon energy $E$, where $l_N$ is the nucleon mean free path and
$|{\rm d}E/{\rm d}x|_c$ is the nucleon energy loss per unit path length due
to 'continuous' energy losses such as multiple Coulomb scattering. When
$E (l_N |{\rm d}E/{\rm d}x|_c)^{-1}\simge 1$ nucleons predominantly loose
energy due to nucleon-nucleon scattering and spallation processes, whereas
in the opposite case nucleons loose their energy predominantly continuously
via multiple electromagnetic scatterings on $e^{\pm}$ and photons 
(cf. Eq. B.1). Only in the former limit nuclear cascades occur.
The quantity is shown for neutrons (green - dashed) at
temperatures $T=90$ and $30\,$keV and protons (red - solid) at temperatures
$T = 30,10,1$ and $0.01\,$keV as labeled in the figure.}
\label{fig_comp_np}
\eef

Hadronically interacting primaries may affect light-element yields in
the whole above quoted decay time range. Charged mesons have an effect only
during early times ($\tau\sim 10^{-1}-10^{1}$sec) due to charge
exchange reactions~\cite{RS} (Rec.34-43 Table~\ref{bigtable}, 
App. E.1) with nucleons, converting mostly
protons to neutrons and thereby increasing the \he4 abundance. At larger
decay
times their effects are negligible since they mostly decay before reacting
with nucleons. At shorter decay
times thermal weak interactions very quickly re-establish the SBBN
neutron-to-proton ratio. Antinucleons may have an effect at all times,
though their effect is the most pronounced during the time interval 
$\tau\sim 10^{-1}-10^{2}$sec (App. E.2). 
As they are more likely to annihilate on protons they also tend to 
increase the
neutron-to-proton ratio. 
At times $\tau\simge 200\,$sec a significant fraction $Y_p\approx 0.25$
of all baryons are bound into helium nuclei. Annihilation of antinucleons
on helium nuclei may leave \h2,\h3, and \he3 as secondaries~\cite{Balestra}. 
However, this
effect is subdominant as compared to the effects by injected nucleons.

\bef
\showone{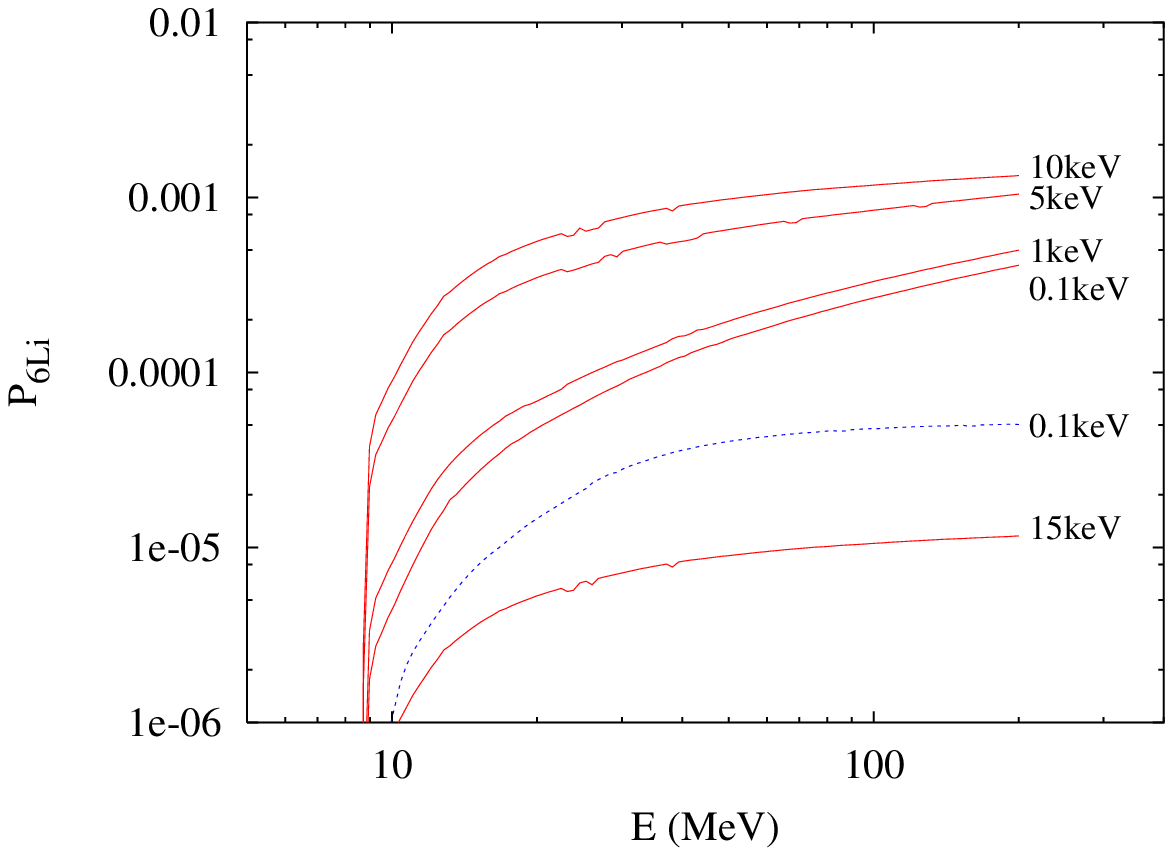}
\caption{Probability $P_{^6{\rm Li}}$
of energetic \h3 nuclei (red - solid) and energetic
\he3 (blue - dotted) of initial energy $E$ to fuse on ambient
\he4 nuclei to form \li6 via the
reactions $\H3(\alpha ,n)\Li6$ and $\He3(\alpha ,p)\Li6$, respectively. 
The figure shows this probability for \h3-nuclei at temperatures
$T = 5,10,5,1$ and $0.1\,$keV and for \he3-nuclei at temperature 
$T = 0.1\,$keV, respectively. Survival of the freshly formed \li6 nuclei
against thermal nuclear reactions is also taken into account as evident
by the comparatively low $P_{^6{\rm Li}}$ at $T = 15\,$keV. The figure
illustrates that $P_{^6{\rm Li}}$ dramatically increases at $T\sim 5-10\,$keV
due to a decrease of the efficiency of Coulomb stopping in this narrow
temperature interval (see text).}
\label{fig_6Li}
\eef

Neutrons at $\tau\simge 200\,$sec and protons at $\tau\simge 10^4$sec
thermalize to a substantial degree via nucleon-nucleon 
collisions (Rec. 48-73 Table~\ref{bigtable}, App.E.3,E.4) and
nuclear spallation reactions 
(Rec. 74-83 Table~\ref{bigtable}, App.E.5). This is because the competing
processes for thermalization, magnetic moment scattering of neutrons
off plasma $e^{\pm}$
(Rec. 3, App. D.1.c), 
Coulomb stopping of protons by plasma $e^{\pm}$ (Rec.1, App. D.1.a), 
and Thomson scattering of protons on CMBR photons (Rec. 2, App. D.1.b)
loose their efficiency. This holds true 
except for very energetic nucleons as well as protons in
the tens of MeV range and is 
due to ever decreasing numbers of $e^{\pm}$ and energy of CMBR photons.
These results may be observed in Fig.~\ref{fig_comp_np}. 
Each nucleon-nucleon scattering produces another energetic nucleon. The
injection of an energetic $100\,$GeV nucleon may therefore lead to
the production of several tens 
of $1\,$GeV nucleons. Their collective impact on the light-element yields
via spallation of \he4 to produce \h2,\h3, and \he3, is far greater than
that of a $100\,$GeV injected antinucleon~\cite{Dimo}. 
The latter undergoes maximal
2-3 scattering before annihilating, with the respective probability of
spallation of \he4 being compartively small. The injection of nucleons is
therefore always important. At early times this is 
due to an increase of the neutron density
($\tau\sim 1-10^4$sec) leading to either an increased 
\he4 ($\tau\simle 200\,$ sec), increased \h2 ($\tau\sim 200-10^4$sec),
or decreased \li7 ($\tau\sim 10^3$sec) abundance. Somewhat later, and
for the whole range of decay times considered here 
($\tau\simge 200\,$sec) spallation of \he4 by energetic
nucleons and their secondaries yields large amounts
of \h2,\h3, and \he3. These trends may be seen in Fig.~\ref{yields}. 
As \he4 spallation processes by neutrons in the time
range $\tau\sim 200\, - 10^4$sec are by far more
efficient that those by protons, inelastic nucleon-nucleon reactions
(Rec. 52-73, App.E.4) 
which preferentially convert protons to neutrons are
also important for accurate BBN yield predictions.

\bef
\showone{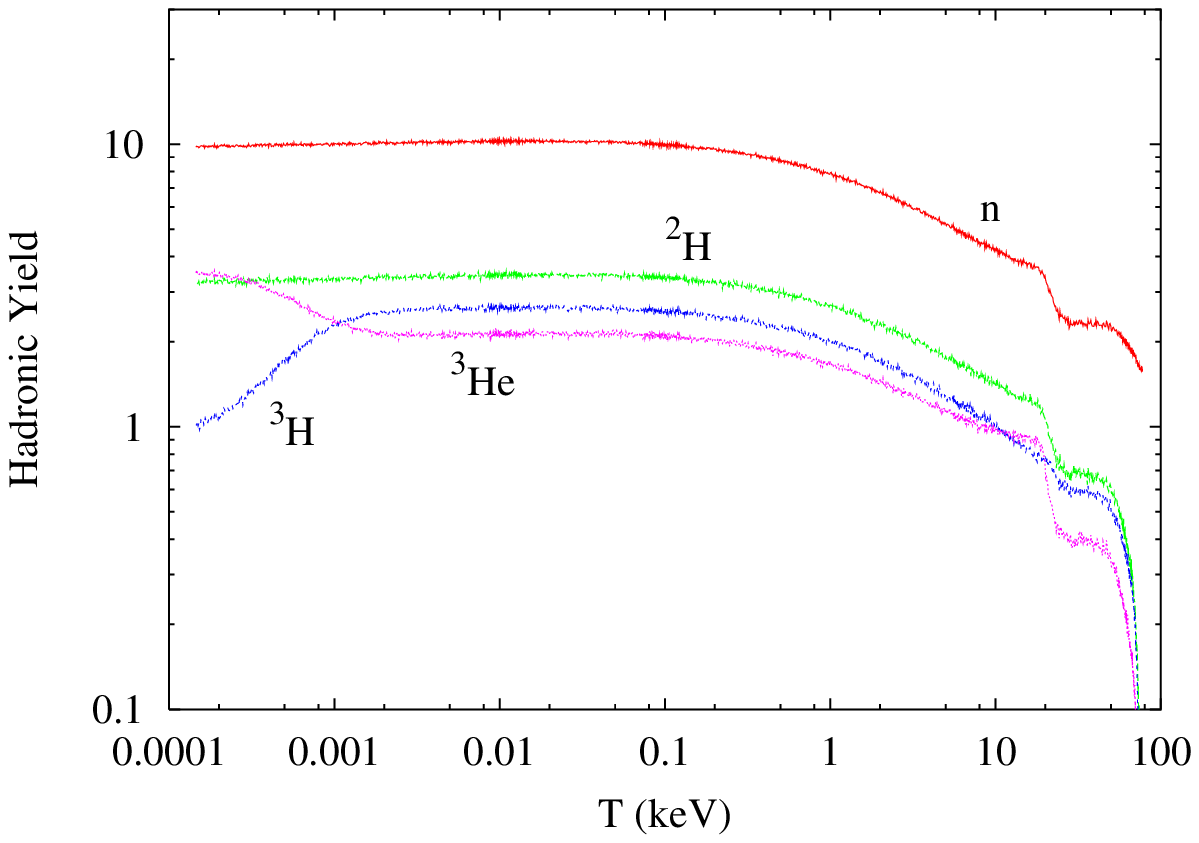}
\caption{Yields of (from top to bottom) neutrons (red), deuterium nuclei
(green), tritium nuclei (blue), and \he3-nuclei (magenta) as a function
of cosmic temperature $T$ per hadronically decaying particle 
$X\to q\bar{q}$ of mass
$M_x = 1\,$TeV, where $q$ denotes a quark.  
Note that initially after hadronization 
of the $q$-$\bar{q}$ state on average only
$1.56$ neutrons result. The remainder of the created neutrons at
lower temperatures $T\simle 90\,$keV are resulting from the thermalization
of the injected neutrons (and protons) due to inelastic nucleon-nucleon
scattering processes and \he4 spallation processes. Similarly, all the
\h2,\h3, and \he3 nuclei are due to  \he4 spallation processes and
$np$ nonthermal fusion reactions (for \h2) induced by the thermalization
of the injected energetic nucleons. The figure does note include the
electromagnetic yields (cf. Fig.~\ref{yieldsEM}) due to photodisintegration
which is inevitable even for a hadronic decay since approximately $45\%$ of
the rest mass energy of the decaying particle is converted into energetic
$\gamma$-rays and energetic $e^{\pm}$ after hadronization and pion decays.}
\label{yields}
\eef

The spallation of \he4 to form \h3 and \he3 is also important as it may
lead to the synthesis of \li6~\cite{Dimo} (cf. App. E.6). 
The mass three nuclei \h3 and \he3 emerge energetic from these
reactions with typical energies around $10\,$MeV. They may thus participate
in the non-thermal fusions reactions 85-86 to form \li6. The formation
of \li7 by those energetic 3-nuclei, as well as by energetic \he4 resulting
from elastic nucleon-\he4 collisions (Rec. 87-88)
is by far less important. Energetic
mass three nuclei 
to the largest part loose their energy efficiently via Coulomb
stopping. However, even the comparatively small fraction 
$\sim 10^{-4}$ (cf. Fig.~\ref{fig_6Li}) which
fuses may yield an observationally important \li6 abundances. The freshly
fused \li6 may survive thermal nuclear reactions only above 
$\tau\simge 10^4$sec. A particularity in the Coulomb stopping 
(cf. App.D.1.a)
which renders it very much less efficient for nuclei with velocities below
the thermal electron velocity, which is the case for $\sim 10\,$ MeV 3-nuclei
at $\tau\sim 10^4$sec, increases the \li6 production efficiency around this time
by a factor of ten. As explained in App. E.6  \li6 yields due to
energetic mass three 
nuclei produced by \he4 spallation are somewhat uncertain and
may well be some tens of per cents higher. 
The uncertainty is due to an incomplete
knowledge of the high-energy tail of produced mass three nuclei which may well
include a backward scattering peak. The present study is therefore
conservative in estimating \li6 yields.

Electromagnetically interacting particles may impact BBN yields only at
comparatively large decay times ($\tau\simge 10^5$sec). This is due to
energetic photons and $e^{\pm}$ interacting efficiently with CMBR 
photons (cf. App. D.2 ).
$\gamma$-rays rapidly pair-produce on CMBR photons (Rec. 4) and electrons 
inverse Compton scatter
(Rec. 5). Comparatively long-lived against further interactions are only
$\gamma$-rays of energies below the $e^{\pm}$ pair production threshold.
This threshold is at $E\approx 2.2\,$MeV at $\tau\sim 10^5$sec and increases
continuously as the CMBR temperature drops reaching 
$E\approx 19.8\,$MeV at $\tau\sim 10^7$sec. These latter energies are the
thresholds for photodisintegration of \h2 and \he4, respectively. Once
$\gamma$-rays have dropped below the pair production threshold their
interactions close to the threshold are dominated by scattering of
CMBR photons (App. D.2.d), and further below the threshold by Bether-Heitler
pair production (App. D.2.a) 
and Compton scattering (App. D.2.c). A substantial though
small fraction $f\sim 0.01$ may photodisintegrate either 
\he4 or \h2~\cite{Lindley}.
The reader is referred to Fig.~\ref{yieldsEM} and Fig.~\ref{destEM} for
typical light-element production- and destruction- factors during
electromagnetic decays.
When \he4 photodisinthegration becomes important ($\tau\simge 10^7$sec),
photodisintegration of \h2 is a subdominant effect as more \h2 is produced
in the relatively larger numbers of \he4-nuclei which are 
photodisintegrated. Energetic \h3 and \he3 produced in large numbers
during the \he4 photodisintegration may lead via non-thermal fusion
reactions to \li6~\cite{J00}, 
similar to the energetic mass three nuclei produced in \he4
spallation during hadronic decays. To a less important degree \li6 may
also be created by direct photodisintegration of \li7 and \be7.

\bef
\showone{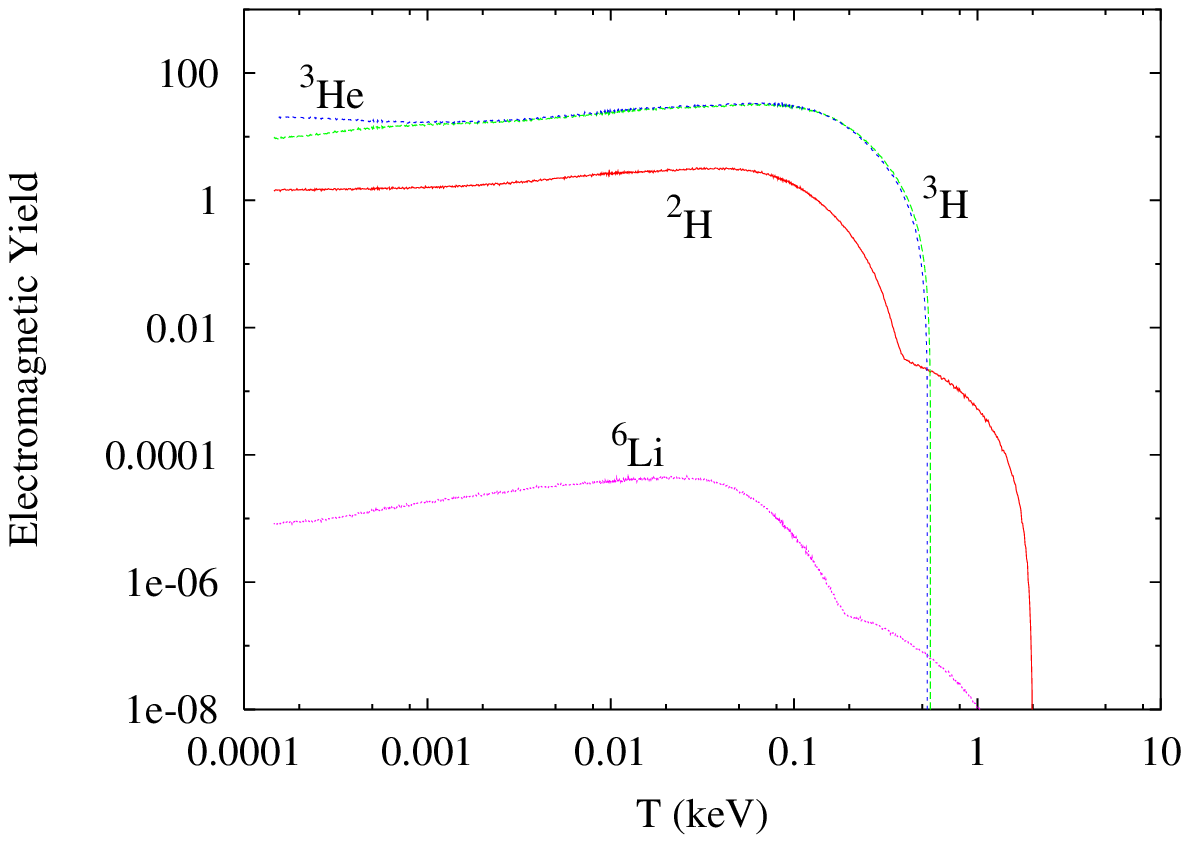}
\caption{Yields of (from top to bottom) \he3 (blue), \h3 (green), 
\h2 (red), and \li6 (magenta) nuclei per TeV of electromagnetically
interacting energy injected
(in form of energetic $\gamma$-rays and energetic $e^{\pm}$)
due to photodisintegration reactions (\h2, \h3, \he3, and \li6) and
fusion reactions (\li6) as a function of cosmic temperature.}
\label{yieldsEM}
\eef

\bef
\showone{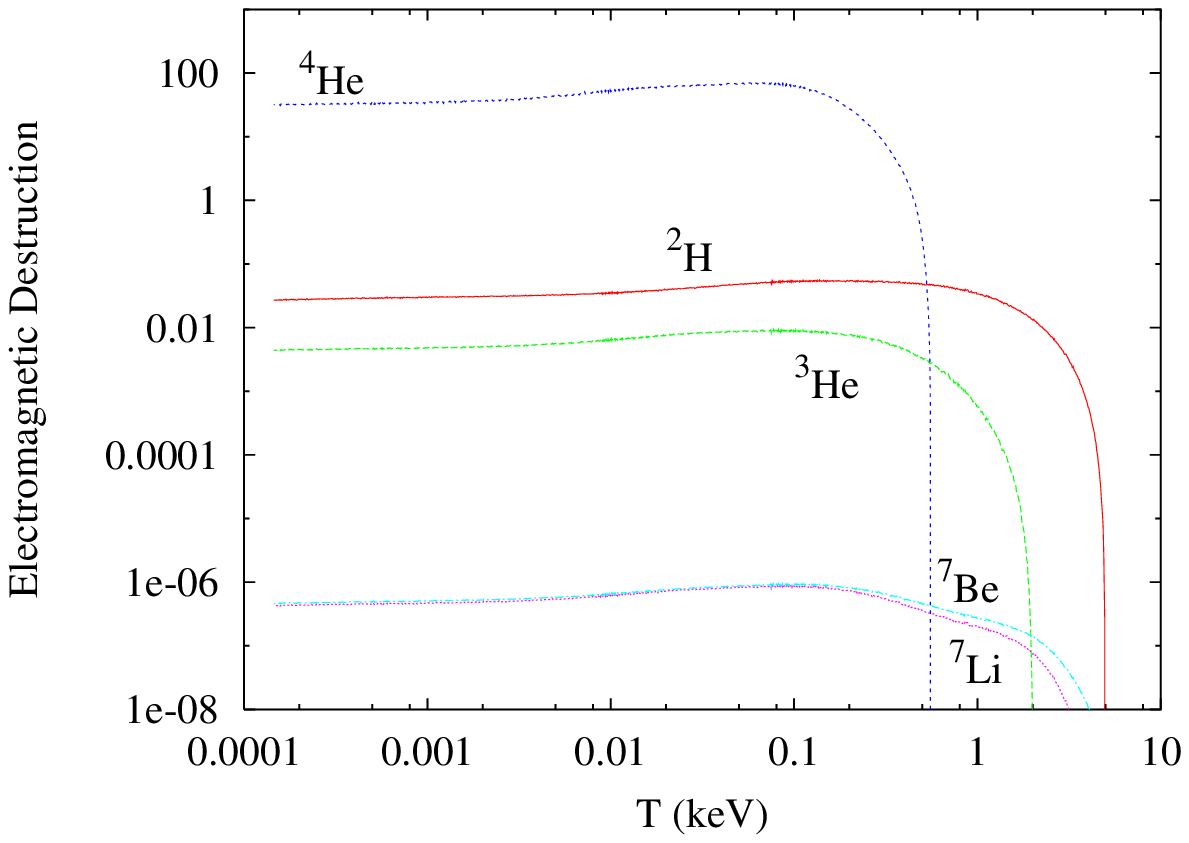}
\caption{Number of destroyed (from top to bottom)
\he4 (blue), \h2 (red), \he3 (green), and \li7 (magenta) (\be7 - light blue)
nuclei per TeV of electromagnetically interacting energy injected into
the primordial plasma at temperature $T$. For the purpose of
illustration we have taken the \li7 and \be7 abundances equal at
\li7/H = \be7/H $\approx 4.34\times 10^{-10}$. In reality it is the sum of both
isotopes which is synthesized at 
(\li7 + \be7)/H $\approx  4.34\times 10^{-10}$ in a SBBN scenario at
$\Omega_bh^2\approx 0.02233$ with \be7 being converted to \li7 by electron
capture at cosmic temperatures $T\approx 0.1 - 1\,$keV.}
\label{destEM}
\eef

Thus, the effects of injection of electromagnetically interacting
particles on BBN yields are due to photodisintegration and are only operative
at comparatively late times ($\tau\simge 10^5$sec). A much simplifying
aspect of BBN with electromagnetically decaying particles is the fact that
results almost always only depend on the total amount of electromagnetically
interacting energy, rather than the energy and nature of the injected
primary. This is due to the number of cascade photons and electrons being
so large that asymptotically the same state is reached, independent of the
initial state. Finally it is noted that hadronic decays (such as 
$X\to q\bar{q}$, with $q$ a quark) generically also lead to the injection
of electromagnetically interacting primaries, as for example due to 
$\pi^0$'s which decay into two photons. 

\section{Observational Limits on the Primordial Abundances} 

The following provides a compilation of the observational
limits on primordial abundances adopted in the present analysis.
For more detail on observations and interpretations, 
the reader is either referred to the original 
literature or reviews on BBN nucleosynthesis (cf. ~\cite{Steigman,BBNreview}).

\subsection{\he4}

The primordial \he4 abundance is inferred from observations of hydrogen- 
and helium- emission lines in extragalactic low-metallicity HII-regions.
The two most recent observers determinations yield 
$Y_p\approx 0.2421\pm 0.0021$~\cite{IT} and 
$Y_p\approx 0.239\pm 0.002$~\cite{Luridiana}. Nevertheless, somewhat smaller 
$Y_p\approx 0.2345$ or
larger $Y_p\approx 0.2443$
determinations by the same authors, but determined from
(partially) different data sets of, have also been cited. The estimates are
significantly below the SBBN $Y_p$ prediction at the WMAPIII baryon
density.
However, the quoted error bars are purely statistical in nature and
only some of the many possible systematic errors could be estimated.
Systematic errors include underlying stellar absorption, ionization
corrections, temperature fluctuations, and collisional excitation.
Recent re-analysis of the original data by an independent group~\cite{OS},
has led to a significantly revised $Y_p\approx 0.249\pm 0.009$ estimate,
including a substantial increase of the error bars. Nevertheless, even this
latter study does not include a study of the majority of systematic
uncertainties, such that its utility is somewhat questionable.
Many of these systematic uncertainties  
would lead again to a decrease of the inferred
$Y_p$. In a conservative spirit, however, the present analysis utilizes
the upper limit as quoted above
\begin{equation}
Y_p < 0.258\,
\label{he4} .
\end{equation} 
A lower limit on $Y_p$ is not considered as usually of no use for
deriving limits on relic decaying particles.

\subsection{\h2}

The most accurate method for an observational determination of the
primordial \h2/H-ratio is believed to be due to observations of
low-metallicity quasar absorption line systems (QALS). Such gaseous
systems, on the line of sight between a high redshift quasar and the
observer,
produce absorption lines in the continuous spectrum of the quasar at
the redshifted wavelength of the Lyman-$\alpha$ \h2 and \h1 positions,
respectively. If the QALS has a simple velocity structure, 
the depths of the absorption troughs may be used to infer the
\h2/H-ratio. As observations focus on systems at very low metallicity, the
assumption is typically that only very little stellar \h2 destruction
has occurred, and that the determined abundance may be directly compared
to the predicted BBN \h2/H-ratio. 
When the currently six best determinations~\cite{D/H}
are averaged one obtains a QALS-inferred primordial \h2/H abundance of
$\H2/{\rm H}\approx 2.4\pm 0.4\times 10^{-5}$~\cite{Steigman}. 
This value compares favorably
to that predicted by a SBBN scenario at the WMAPIII baryonic density.
Nevertheless, caution has to be applied as the intrinsic dispersion between
individual central values of \h2/H determinations is much larger than
most of the quoted error bars for individual QALS. Thus,
the highest individual
2-$\sigma$ upper limit and the lowest individual
2-$\sigma$ lower limit of all six
systems span a range
\begin{equation}
 1.2\times 10^{-5}\simle\H2/{\rm H}\simle 5.3\times 10^{-5}\, ,
\label{h2}
\end{equation}
much larger than the above quoted error bar. Furthermore, a systematic
trend of decreasing \h2/H with increasing column density $N_H$ of the QALS may
be found indicating the existence of possible systematic errors. 
Attempting to remain on the conservative side the above range is adopted
in the analysis for an acceptable \h2/H abundance. Conservatism may be
in order not only because a recent proposal of a very active stellar
Pop III population, which could potentially reduce the primordial \h2/H-ratio 
in certain (high $N_H$) systems without much trace of associated 
iron or silicon production~\cite{Piau}.
 
\subsection{\he3/\h2}

Precise observational determinations of \he3/H-ratios are only possible
within our galaxy. Since our galaxy is chemically evolved, it is difficult
to make a straightforward connection to the primordial \he3/H abundance.
Since \he3, in fact, is known to be destroyed in some stars, but produced
in others, observed \he3/H-ratios may also only with difficulty
be utilized to obtain an upper or lower limit on the primordial \he3/H.
This is in contrast to the \he3/\h2-ratio. \h2 is known to be
always destroyed in stars (in fact is burned into \he3), whereas \he3 is
either destroyed or produced. The cosmic \he3/\h2-ratio may therefore
only increase with time. Thus a determination of this ratio in any
environment, even chemically evolved, may be taken as an upper limit on the
primordial \he3/\h2-ratio~\cite{S95}. 
From a combination of solar wind observations
and the planetary gas component of meteorites and Jupiter, it is possible
to infer the \he3/H$\approx 1.66\pm 0.05\times 10^{-5}$ and
\h2/H$\approx 1.94\pm 0.39\times 10^{-5}$-ratios at the time of solar system
formation independently~\cite{GG}. Employing the 2-$\sigma$ upper end of
the \he3 abundance and the 2-$\sigma$ lower end of the \h2 abundance, when
may derive a conservative constraint of
\begin{equation}
\He3/\H2 < 1.52\, ,
\label{he3h2}
\end{equation}
for the primordial \he3/\h2 ratio.

\bef
\showone{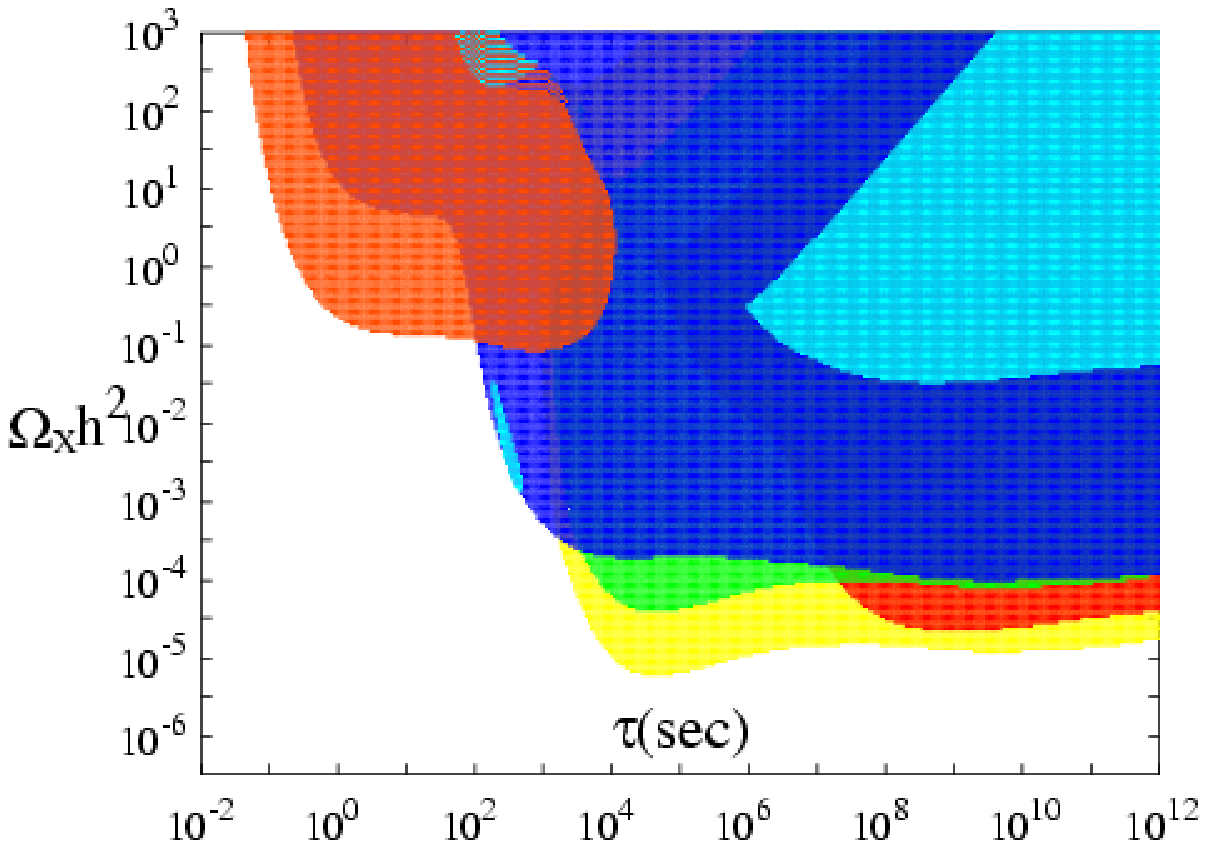}
\caption{Conservative
BBN constraints on the abundance of relic decaying neutral particles as
a function of their life time for a $M_x = 1\,$TeV particle with hadronic
branching ratio $B_h = 1$. Limits are given on the contribution 
the decaying particles would have made to the present critical density,
$\Omega_X h^2$ (with $h$ the Hubble parameter), if they would have not decayed.
For a conversion to constraints on, for example $M_X n_X/n_{\gamma}$ 
the reader is referred to App. A.
The colored regions are excluded and correspond to the constraints imposed by
the observationally inferred upper limit on \he4 - orange - (Eq.~\ref{he4}), 
upper limit on \h2 - blue - (Eq.~\ref{h2}), upper limit on
\he3/\h2 - red - (Eq.~\ref{he3h2}),
and lower limit on
\li7 - light blue - (Eq.~\ref{li7}). Conservative constraints derived from
\li6/\li7 (Eq.~\ref{672})  are shown by the green region. 
The region indicated by yellow violates
the less conservative \li6/\li7 (Eq.~\ref{671})
constraint but should not be considered ruled out. Rather,
this region may
be cosmologically interesting as a putative source of \li6 
in low-metallicity stars by relic decaying particles.}
\label{BH=1E0M=1TeV}
\eef

\subsection{\li7}

For a long time it has been known that the \li7/H-ratio in the atmospheres
of low-metallicity 
PopII stars is essentially constant with metallicity (the Spite
plateau) with deduced central values on the plateau falling in
the range 
$1.23\times 10^{-10}\simle \Li7/{\rm H}\simle 1.73\times 10^{-10}$~\cite{BM}. 
As this value is traditionally interpreted to be close to the
primordial \li7/H-ratio, a factor $2-3$ discrepancy between the SBBN
predicted primordial \li7/H at the WMAPIII baryon density (usually in
excess of $4\times 10^{-10}$ and that inferred from the stars on the
Spite plateau is apparent. This situation has neither been changed by a
large number of re-evaluations of the \li7/H-abundance of the Spite plateau,
nor convincingly shown to be solved by stellar depletion of \li7.
As the \li7 isotope is not very important in constraining relic
decaying particles (rather, it may resolve the problem), 
the reader is referred elsewhere for discussion of
this problem (e.g.~\cite{Steigman,Jetal05,Piau}). 
The present analysis adopts only
a lower limit on the primordial \li7/H
\begin{equation}
\Li7/{\rm H} > 0.85\times 10^{-10}\, ,
\label{li7}
\end{equation}
chosen by the 95\% confidence level lower limit as derived 
by Ref.~\cite{Ryan}. 

\bef
\showone{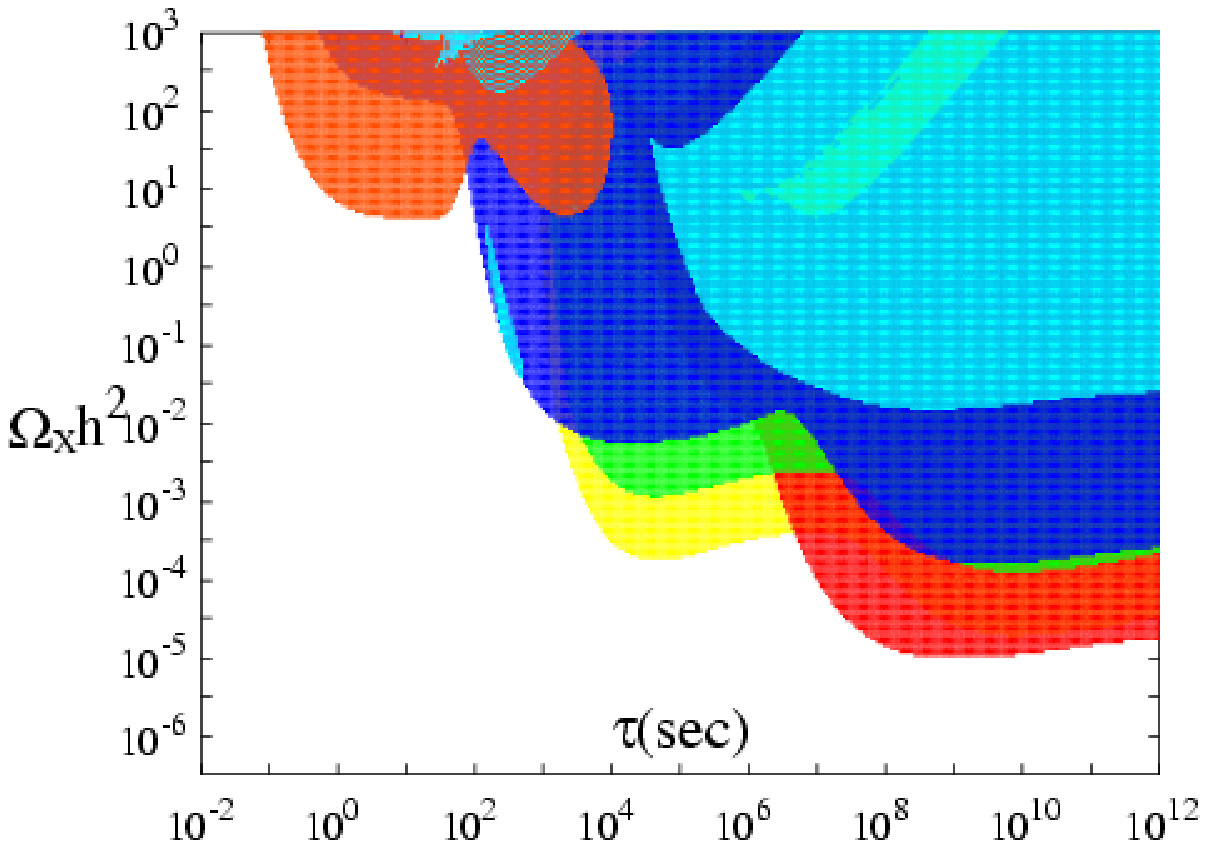}
\caption{As Fig.~\ref{BH=1E0M=1TeV} but for hadronic branching ratio
$B_h = 3.333\times 10^{-2}$.}
\label{BH=3E-2M=1TeV}
\eef

\subsection{\li6}

The isotope of \li6 can be particularly useful in deriving constraints
on relic decaying particles~\cite{J00}. Recently, the number of claimed
preliminary detections of \li6/\li7-ratios in low-metallicity Pop II stars
has multiplied by a large factor~\cite{Asplund} (cf. to Ref.~\cite{6Li} for
the few former detections) falling in the range 
$0.03\simle\Li6/\Li7\simle 0.07$ with average \li6/\li7$\approx 0.042$ and
with no star having \li6/\li7 in excess of $0.1$. 
The origin of this \li6
is unknown. Concerning a limit on the primordial \li6-abundance the situation
is somewhat problematic. This is particularly so, since the predicted SBBN
\li7 is in excess of that observed. If this latter fact is after
all explained by stellar \li7 destruction, \li6 would also be destroyed. 
In the best case (destroying as
little as possible \li6), both the \li6 and \li7 abundances would be destroyed
by the same factor, with their ratio, nevertheless, staying constant. In this
case one may adopt
\begin{equation}
\Li6/\Li7\simle 0.1
\label{671}
\end{equation}
as upper limit. This would correspond to a primordial \li6 abundance
$\Li6/{\rm H}\simle 4.34\times 10^{-11}$ 
(given that the SBBN prediction is $4.34\times 10^{-10}$ in the present
analysis, cf. App.F)
already much in excess of the
2-$\sigma$ upper limit of \li6/H$\, < 1.47\times 10^{-11}$ in the
well studied star HD84937. However, calculations of \li6 and \li7 destruction
in stars generally predict more \li6 destruction than \li7 destruction,
since \li6 is more fragile than \li7.
In Ref.~\cite{Pinn} 
which studied \li7 depletion in rotating stars a correlation
between the amount of \li7 and \li6 destruction was obtained, with, for
example, factor
$2\, (4)$ reduction of the \li6/\li7 ratio occurred for factor $1.6\, (2.5)$
reduction of the \li7 abundance.
When the (fine tuned) models by Ref.~\cite{Richards} are
taken, designed to explain a larger amount of \li7 destruction without
spoiling the flatness of the Spite plateau (when plotted against stellar
temperature), a factor $1.6\, (2.5)$ of \li7 destruction implies a factor 
$1.25\, (15.8)$ reduction in the \li6/\li7-ratio. Particularly the last
destruction factor would render the \li6 isotope not anymore as useful
in deriving limits on decaying particles. Nevertheless, in the spirit of a
conservative study we will also adopt
\begin{equation}
\Li6/\Li7\simle 0.66 \quad\quad {\rm conservative}\, ,
\label{672}
\end{equation}
as a second more conservative limit. 
Here this value derives by taking the average observed \li6/\li7 ratio
of $0.042$
and multiplying it by the \li6/\li7 destruction factor of $15.8$.
This is done with the understanding that any models 
which violate Eq.~(\ref{671}) but not Eq.~(\ref{672}) should be flagged
as potentially being ruled out. On the other hand, such models,
if not violating other observational constraints, could also explain the
origin of the observed \li6 (cf. Ref.~\cite{J04-01,Jetal05}).

\bef
\showone{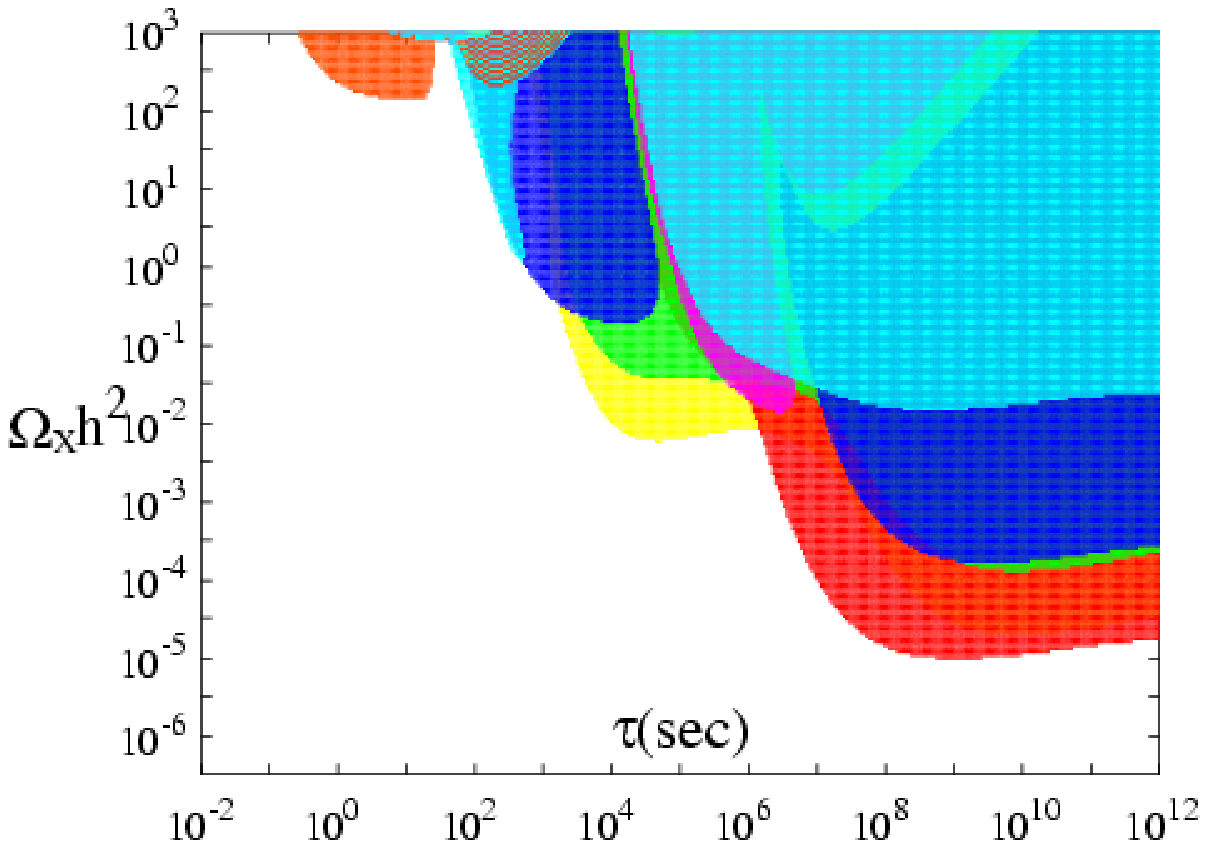}
\caption{As Fig.~\ref{BH=1E0M=1TeV} but for hadronic branching ratio
$B_h = 10^{-3}$. The region excluded by the lower limit on \h2/H 
(Eq.~\ref{h2}) is indicated by the color magenta.}
\label{BH=1E-3M=1TeV}
\eef

\bef
\showone{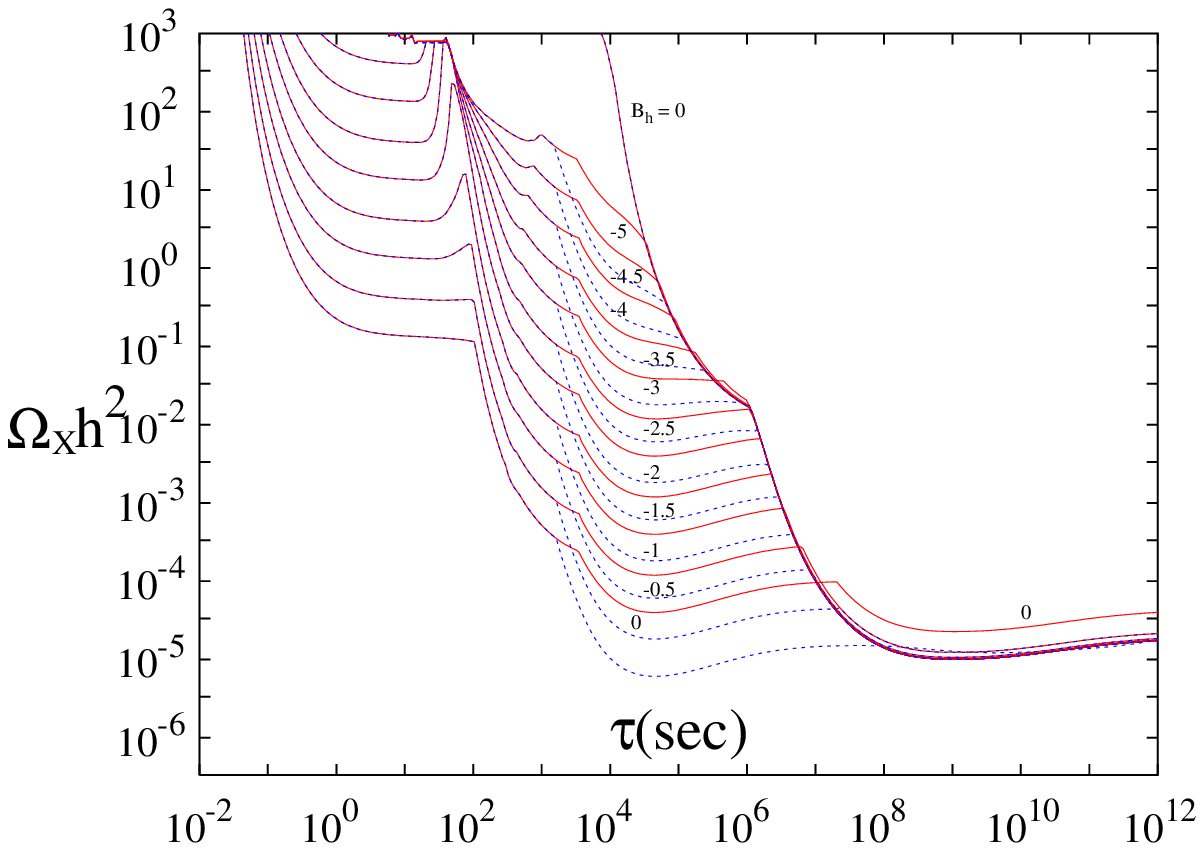}
\caption{Constraints on the abundance of decaying neutral particles
as a function of life time for an $M_X = 1\,$TeV particle with varying
hadronic branching ratios. Constraints are shown for hadronic branching
ratios (from bottom to top) ${\rm log}_{10} B_h = 0$, $-0.5$, $-1$, -1.5,
-2, -2.5, -3, -3.5, -4, -4.5, and -5, respectively, as labeled. The case
$B_h = 0$ is also shown. The labels always correspond to the solid (red) line
above it. For each $B_h$ the region above this
solid (red) line is ruled out when conservative constraints are applied
(i.e. Eqs~\ref{he4}-\ref{li7} and \ref{672}). When the \li6/\li7 constraint 
Eq.~\ref{672} is replaced by the less conservative Eq.~\ref{671} the dotted
(blue) constraint lines result. These dotted lines
coincide for small and large $\tau_X$
with the solid (red) lines.}
\label{BH1}
\eef

\bef
\showone{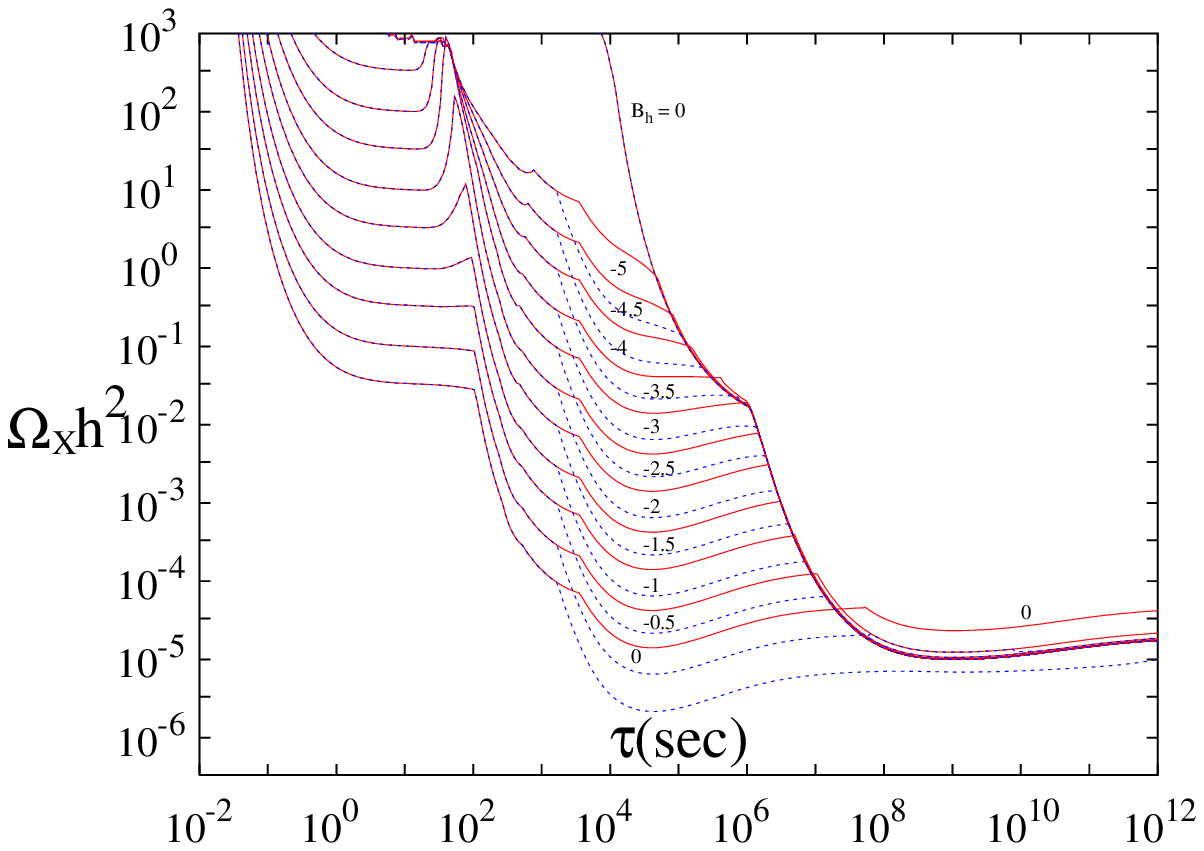}
\caption{As Fig.~\ref{BH1} but for $M_X = 100\,$GeV.}
\label{BH2}
\eef

\section{Constraints on Relic Decaying Particles}

In this section conservative constraints on the abundance of putative
relic decaying neutral particles in the early Universe are presented. The number
density $n_X$ of an early produced semi-stable particle species $X$ with
decay width $\Gamma_X = 1/\tau_X$ follows the equation
\begin{equation}
\frac{{\rm d} n_X}{{\rm d} t} = - 3 H n_X -\frac{n_X}{\tau_X}\, ,
\end{equation}
where $t$ is cosmic time and $H$ is the Hubble constant. Here the first
term on the right-hand-side describes dilution due to cosmic expansion, 
whereas the second term represents the particle decay. Results are presented
for a large number of different hadronic branching ratios $B_h$ of the
particle $X$. In particular, it is assumed that a fraction $B_h$ of all
decays are hadronic and a fraction $1-B_h$ are electromagnetic. During
hadronic decays the primaries injected into the plasma are assumed to be
a quark-antiquark pair of total energy $M_X$ and total momentum zero in the
cosmic rest frame. The hadronization of the $q$-$\bar{q}$ fluxtube is
followed with aid of the code PYTHIA, resulting in numerous nucleons, mesons,
$e^{\pm}$ and $\gamma$-rays injected into the plasma. Note that for large
$M_X$, as considered here, the spectrum of the resulting post-hadronization 
particles is virtually independent of the type of initial quark injected.
As already stated in Sec. II, for sufficiently massive particles $X$
the effects of electromagnetic decays on BBN yields depends only on the
total energy $E_{e,\gamma}$ in $e^{\pm}$ and $\gamma$-rays 
injected into the plasma. For electromagnetic decays 
$E^{\rm EM}_{e,\gamma} = M_X$ is
thus assumed with no injection of nucleons or mesons. 
Hadronic decays are also associated with the release of energetic 
electromagnetically interacting particles. Following results
from PYTHIA $E^{\rm H}_{e,\gamma} = 0.45 M_X$ is taken.
Finally, it is noted that only baryon number conserving decays are considered.

In Figs~\ref{BH=1E0M=1TeV}-\ref{BH=1E-3M=1TeV} constraints on the abundance
of $M_X = 1\,$TeV relic decaying particles for hadronic branching
ratios $B_h = 1$, $B_h = 3.333\times 10^{-2}$, and $B_h = 10^{-3}$, 
respectively, are shown. The colored (shaded) regions are excluded, with
different colors (shades) corresponding to constraints derived from different
light elements, as indicated in the figure captions. It is evident that
the most stringent constraints derive from an overproduction of \he4 at
early times $\tau_X\simle 10^2$sec, an overproduction of \h2 in the
decay time interval $10^2{\rm sec}\simle \tau_X\simle 10^3{\rm sec}$,
an overproduction of \li6 in the decay time interval 
$10^3{\rm sec}\simle \tau_X\simle 10^7{\rm sec}$, and an overproduction
of the \he3/\h2-ratio for large decay times $\tau_X\simge 10^7$sec.
The yellow area in the figures corresponds to elevated
\li6 production in the range 
$0.1\simle \Li6/\Li7\simle 0.66$ and should currently not be considered
ruled out. Though in excess of the observationally inferred \li6/\li7
abundance ratio in low-metallicity stars it is conceivable that some of the
produced \li6 has been destroyed in low-metallicity stars.The reader is
referred to Sec. III.E for details.
In contrast, as the source of the unexpected large abundance of \li6 in 
low-metallicity stars is unknown, the yellow region may be considered as
cosmologically interesting~\cite{J04-01,Jetal05}. 

In Figs~\ref{BH1}-\ref{BH2} constraints on the abundance of relic decaying
particles for a large number of hadronic branching ratios and two
different particle masses, $M_X=100\,$GeV and $M_X=1\,$TeV are shown. 
The region above the solid (red) lines for each particular $B_h$ are excluded.
The figure also shows by the dotted (blue) lines
how the constraints would change if one were to
replace the observationally inferred limit on \li6/\li7 in
Eq.~\ref{672} by the less conservative constraint
Eq.~\ref{671}. However, as argued above, the latter constraint 
should currently not be applied.

\section{Conclusions}

BBN is a powerful probe of the conditions in the early Universe and
may severely limit the putative existence of relic decaying particles.
In this paper, the effects of
electromagnetically and hadronically decaying
particles on the BBN light-element synthesis of \h2, \he3, \he4, 
\li6, and \li7 have been considered in detail. 
All processes which the author
believes to play a role at the $10\%$ level have been included.
These include nucleon-nucleon elastic and inelastic scatterings, 
nucleon-antinucleon annihilation, spallation of \he4 by energetic nucleons,
fusion reactions of \h3 and \he3 on \he4 nuclei, mesonic charge exchange,
Coulomb- and magnetic- moment scattering, Thomson scattering, Bethe-Heitler
pair production, photon-photon scattering, (inverse) Compton scattering,
and photodisintegration of light nuclei, among others. The analysis attempts
to stay as close as possible to existing experimental data. A number
of expressions, as for example the Coulomb stopping power of comparatively slow
nuclei have been newly evaluated. All these non-thermal reactions are
coupled to the thermonuclear fusion reactions known to be of importance
during BBN. Results of such complete calculations have been first presented
in Ref.~\cite{J04-01,KKMlett} and particularly, Ref.~\cite{KKM}, nevertheless, 
only for a limited number of decaying particle properties.

Predicted light-element yields are compared to 
observationally inferred limits on primordial light-element abundances
to infer accurate and conservative
limits on electromagnetically and hadronically decaying
particles in the early Universe. Constraints on the abundance of such
decaying
particles are shown for a large number of 
hadronic branching ratios and particles masses. This was necessary 
to obtain comparatively accurate limits as a simple extrapolation between
the results of only two hadronic branching ratios may yield to erroneous or
inaccurate
results. The results in this paper may be used to further constrain the
evolution of the very early Universe.


\begin{footnotesize}
\begin{table}
\newcommand{\lstrut}{{$\strut\atop\strut$}}
\caption{}
\label{T:oh2}
\vspace{2mm}
\begin{center}
\begin{tabular}{|| c | l | l | c || c | l | l | c ||}
\hline
No. &  Reaction  & Channel & Ref. & No. &  Reaction  & Channel & Ref.\\
\hline
1 & $p + e^{\pm}\to$ elastic & Coulomb & App. D.1.a & 48 & $p + p\to$ & elastic & ~\cite{Meyer,SAID}, App.E.3\\
2 & $p\, (N) + \gamma_{\rm CMBR}\to$ & Thomson &  App. D.1.b & 49 & $p + \He4\to$ & elastic & ~\cite{Meyer}, App.E.3 \\
3 & $n + e^{\pm}\to$ & magnetic moment & App. D.1.c & 50 & $n + p\to$ & elastic & ~\cite{Meyer,SAID}, App.E.3 \\
4 & $\gamma + \gamma_{\rm CMBR}\to$ & $e^- + e^+$ & App. D.2 & 51 & $n + \He4\to$ & elastic & ~\cite{Meyer}, App.E.3\\
5 & $e^{\pm} + \gamma_{\rm CMBR}\to$ & inverse Compton & App. D.2, D.2.b & 52 & $p + p\to$ & $p + p + \pi^0$  & ~\cite{Bystricky}, App. E.4\\
6 & $\gamma + p ({\rm ^4He})\to$ & $p ({\rm ^4He}) + e^- + e^+$ & App. D.2.a  & 53 & & $n + p + \pi^+$  & ~\cite{Bystricky,remark200}, App. E.4\\
7 & $\gamma + e^-\to$ & Compton & App. D.2.c  & 54 & & $\H2 + \pi^+$ & ~\cite{Bystricky,remark200}, App. E.4\\
8 & $\gamma + \gamma_{\rm CMBR}\to$ & $\gamma +\gamma$ & App. D.2.d  & 55 & & $p + p + \pi^+\pi^-$ & ~\cite{Bystricky}, App. E.4\\
 & & &   & 56 & & $\H2 + \pi^+\pi^0$ & ~\cite{Bystricky}, App. E.4\\
9 & $\gamma + \H2 \to$ & $p + n$ &  ~\cite{Cybphoto} & 57 & & $\H2 + 2\pi^+\pi^-$ & ~\cite{Bystricky}, App. E.4\\
10 & $\G + \H3\to$ & $\H2 + n$ & ~\cite{Cybphoto} & 58 & &$p + p + \pi^+\pi^-\pi^0$  & ~\cite{Bystricky}, App. E.4\\
11 &               & $2n + p$ & ~\cite{Cybphoto} & 59 & & $n + p + 2\pi^+\pi^-$ & ~\cite{Bystricky}, App. E.4\\
12 & $\G + \He3\to$ & $\H2 + p$ & ~\cite{Cybphoto} & 60 & & $p + p + 2\pi^0 + (\pi^0s)$ & ~\cite{Bystricky,remark200}, App. E.4\\
13 &                & $2p + n$ & ~\cite{Cybphoto} & 61 & & $n + n + 2\pi^+ + (\pi^0s)$ & ~\cite{Bystricky}, App. E.4\\
14 & $\G + \He4\to$ & $\H3 + p$ & ~\cite{Cybphoto} & 62 & & $n + p + \pi^+ + (\pi^0s)$ & ~\cite{Bystricky,remark200}, App. E.4\\
15 &                & $\He3 + n$ & ~\cite{Cybphoto,remark97}  & 63 & $n + p\to$ & $p + n + \pi^0 + (\pi^0s)$  & ~\cite{Bystricky}, App. E.4\\
16 &                & $\H2 + \H2$ & ~\cite{Cybphoto} & 64 & & $\H2 + \pi^0 +(\pi^0s)$  & ~\cite{Bystricky}, App. E.4\\
17 &                & $\H2 + n + p$ & ~\cite{Cybphoto} & 65 & & $p + p + \pi^-$ & ~\cite{Bystricky}, App. E.4\\
18 & $\G + \Li6\to$ & $\He4 + n + p$ & ~\cite{Cybphoto,remark98} & 66 & & $n + n + \pi^+$& ~\cite{Bystricky}, App. E.4\\
19 &                & ${\rm ^3A} + X$ & ~\cite{Cybphoto} & 67 & & $n + p + \pi^-\pi^+$  & ~\cite{Bystricky,remark200}, App. E.4\\
20 & $\G + \Li7\to$ & $\He4 + \H3$ & ~\cite{Cybphoto,remark99} & 68 & & $p + p + \pi^-\pi^0 +(\pi^0s)$  & ~\cite{Bystricky}, App. E.4\\
21 &                & $\Li6 + n$ & ~\cite{Cybphoto} & 69 & & $\H2 + \pi^-\pi^+$  & ~\cite{Bystricky}, App. E.4\\
22 &                & $\He4 + 2n + p$ & ~\cite{Cybphoto} & 70 & & $n + n + \pi^+ +(\pi^0s)$ & ~\cite{Bystricky,remark200}, App. E.4\\
23 &                & $\He4 + \H2 + n$ & App. D.3 & 71 & & $n + p + 2\pi^-2\pi^+$ & ~\cite{Bystricky}, App. E.4\\
24 &                & $\He6 + p\to \Li6 + p$ & App. D.3 & 72 & & $n + p + \pi^-\pi^+\pi^0$ & ~\cite{Bystricky}, App. E.4\\
25 &                & $2\H3 + p$ & App. D.3 & 73 & & $p + p + 2\pi^-\pi^+$ & ~\cite{Bystricky}, App. E.4\\
26 &                & $\H3 + \He3 + n$ & App. D.3 & & & & \\
27 & $\G + \Be7\to$ & $\He4 + \He3$ & ~\cite{Cybphoto,remark100} & 74 & $p + \He4\to$ & $\H3 + 2p +(\pi s)$ & App. E.5,E.6\\
28 &                & $\Li6 + p$ & ~\cite{Cybphoto} & 75 & & $\He3 + n + p+(\pi s)$ & App. E.5,E.6 \\
29 &                & $\He4 + 2p + n$ & ~\cite{Cybphoto} & 76 & & $\H2 + 2p + n+(\pi s)$ & App. E.5\\
30 &                & $\He4 + \H2 + p$ & App. D.3 & 77 & & $\He3 + \H2$ & App. E.5,E.6\\
31 &                & $\Be6 + n\to \He4 + 2p + n$ & App. D.3 & 78 & & $\H2 + \H2 + p+(\pi s)$ & App. E.5\\
32 &                & $2\He3 + n$ & App. D.3 & 79 & $n + \He4\to$ & $\He3 + 2n+(\pi s)$ & App. E.5,E.6 \\
33 &                & $\H3 + \He3 + p$ & App. D.3 & 80 & & $\H3 + n + p+(\pi s)$ & App. E.5,E.6 \\
   & & & & 81 & & $\H2 + 2n + p+(\pi s)$ & App. E.5\\
34 & $\pi^+ + n\to$ & $\pi^0(\G) + p$ & \cite{RS}, App. E.1 & 82 & & $\H3 + \H2$ & App. E.5,E.6 \\
35 & $\pi^- + p\to$ & $\pi^0(\G) + n$ & \cite{RS}, App. E.1 & 83 & & $\H2 + \H2 + n+(\pi s)$ & App E.5\\
36 & $K^- + n\to$ & $p + X$ & \cite{RS}, App. E.1 & 84 &$p + \Li6\to$ & $\He3+\He4$ & App. E.5\\
37 &              & $n + X$ & \cite{RS}, App. E.1 & & & & \\
38 & $K^- + p\to$ & $n + X$ & \cite{RS}, App. E.1 & 85 & $\H3 + \He4\to$ & $\Li6 + n$ & App. E.6\\
39 &              & $p + X$ & \cite{RS}, App. E.1 & 86 & $\He3 + \He4\to$ & $\Li6 + p$ & App. E.6\\
40 & $K_L + n\to$ & $p + X$ & \cite{RS}, App. E.1 & 87 & $\He4 + \He4\to$ & $\Li6 + \H2\, (\Li6 + p+n$) & App. E.6\\
41 &              & $n + X$ & \cite{RS}, App. E.1 & 88 & & $\Li7 + p\, (\Be7 + n)$ & App. E.6 \\
42 & $K_L + p\to$ & $n + X$ & \cite{RS}, App. E.1 & & & & \\
43 &              & $p + X$ & \cite{RS}, App. E.1 & 89 & $\He3 + p\to$ & elastic & App. E.6\\
&  & & & 90 & $\H3 + p\to$ & $\He3 + n$ & App.E.6 \\ 
44 & $\bar{p} + p\to$ & $\pi$'s & \cite{RS}, App. E.2 & 91 & $\H3 + p\to$ & elastic  & App. E.6 \\
45 & $\bar{n} + p\to$ & $\pi$'s & \cite{RS}, App. E.2 & 92 & $\H3\to$ & $\He3 + e^{-} +\bar{\nu_e}$ & ~\cite{particle}, App.E.6\\
46 & $\bar{p} + n\to$ & $\pi$'s & \cite{RS}, App. E.2 & 93 & $\Be7 + e^-\to$ & $\Li7 + \nu_e$ & ~\cite{particle}\\
47 & $\bar{n} + n\to$ & $\pi$'s & \cite{RS}, App. E.2 & & & & \\
\hline 
\hline
\label{bigtable}
\end{tabular}
\end{center}
\end{table}
\end{footnotesize}

\acknowledgments
I acknowledge useful exchanges with D. Arndt, M. Asplund, A. Blinov, J. Bystricky, R. Cyburt, 
M. Chadeyeva
P. Descouvemont, E. Keihanen, H.-P. Morsch, K. Olive,
N. Prantzos, O. Richard,
E. Vangioni-Flam.

\appendix

\section{General Considerations}

For the computation of light-element abundance yields in the
presence of decaying or annihilating relic particles during
or after BBN one has to essentially treat the thermalization
process of energetic nucleons, antinucleons, and $\gamma$-rays. 
These latter particles result as primaries from the decay/annihilation
process. Their thermalization occurs due to a 
host of processes discussed in detail in App. D and E. At the same time,
thermal nuclear reactions of the thermal baryonic component
may play a role at high temperatures $T\simge 10\,$keV. 
In general, the problem simplifies considerably due to two facts: (a)
thermalization may be treated as an instantaneous process on the
Hubble time scale and (b) interactions between two non-thermal nucleons/nuclei
are improbable. Given a typical proton density 
$n_p\approx 1.6\times 10^{16}{\rm /cm^3}(T/{\rm 10\, keV})^3$ and a
typical nucleon (nuclei)-nucleon interaction cross section of $30\,$mb
and $300\,$mb
at nucleon (nuclei) kinetic energy $E_k\approx 1\,$GeV and $10\,$MeV,
respectively, one finds a typical scattering time 
$\tau_N\approx 0.07{\rm sec\,}(T/{\rm 10\, keV})^{-3}$ 
which is approximately constant with kinetic
energy. This may be compared to the Hubble time 
$\tau_H\approx 1.4\times 10^4(T/{\rm 10\, keV})^{-2}$sec 
to yield a ratio $10\tau_N/\tau_H\approx
5\times 10^{-5}(T/{\rm 10\, keV})^{-1}$ where it is approximated that it
takes around ten scatterings for the complete thermalization of a
nucleon. It is seen that only at very low temperatures $T\simle 0.1\,$eV
thermalization may not be regarded instantaneous on the Hubble scale.
Due to the short survival times of energetic nucleons (nuclei) only
for excessively large
ratios of non-thermalized-to-thermalised nucleons 
$\Omega_{\rm non-thermal}/\Omega_b\sim 10^4(T/{\rm 10\, keV})$ 
interactions between two non-thermalised
nucleons become important. For such large $\Omega_{\rm non-thermal}$,
however, BBN (almost) always fails badly in predicting the observationally
inferred primordial abundances, such that such interactions may be
neglected. 

Abundance yields due to decaying particles may be often understood as the
product of individual factors.
For example, for the production of \li6 nuclei by the fusion reactions
85 and 86, induced by non-thermal mass-3 nuclei, one may
schematically write the final number of produced 
\li6 nuclei $\Delta N_{\Li6}$, produced during a certain cosmic
expansion interval $\rm d\, ln\,a$
\begin{equation} 
\Delta N_{\Li6} = \int {\rm d}N_{\Li6} = \int{\rm d\, ln}a
\biggl(\frac{\rm d\, Decays}{{\rm d\, ln}R}\biggr)
\biggl(\frac{n_{p,n}}{\rm Decay}\biggr)
\biggl(\frac{n_{3}}{n_{p,n}}\biggr)
\biggl(\frac{n_{6}}{n_{3}}\biggr)\, ,
\label{A1}
\end{equation}
where $a$ denotes the cosmic scale factor. Here the integrand is composed
of the number of relic particle
decays per logarithmic scale factor interval, times
the number of energetic nucleons produced per decay, times the number
of energetic \h3 and \he3 produced during the thermalization of a nucleon,
times the fraction of \h3 and \he3 nuclei which fuse to form \li6.
Similarly there are contributions from photodisintegration not shown
in Eq.~\ref{A1}.
The present analysis treats in detail the third and fourth factors, 
pertaining to thermalization.
For example, the last factor in the integrand
$(n_6/n_3)$ may be seen in Fig.~\ref{fig_6Li} for various
cosmic temperatures. 
The second factor is given by
an assumed initial post decay state (e.g. $q\bar{q}$ an energetic 
quark-antiquark fluxtube at an assumed energy) in combination
with the results of an hadronization code, such as PYTHIA,
which computes the number and energy of produced (anti)nucleons, mesons,
and $\gamma$-rays after hadronization of the initial $q\bar{q}$-fluxtube. 
This factor $(n_{p,n}/{\rm Decay})$ convolved with 
$(n_3/n_{p,n})$ may be seen in Fig.~\ref{yields}
(cf. also to Fig.~\ref{yieldsEM} for a similar convolution due to
photodisintegration).
Finally, the first factor is given by the assumed
abundance and life time of the putative relic. 

All figures in the main text limit 
$\Omega_Xh^2$, the contribution of the decaying particle density to
the present critical density $\Omega_X$ if the particle were not decaying. Here
$h$ is the present Hubble constant in units of $100\,$km s$^{-1}$Mpc$^{-1}$.
Other studies often present limits on the number of X-particles per photon
times $M_X$, the mass of the X-particle, or the number of X-particles
per radiation entropy times $M_X$. The conversion factors between
these quantities are given by 
$\xi_1 = n_XM_X/n_{\gamma} = 2.5784\times 10^{-8}\,{\rm GeV}\Omega_Xh ^2$
and
$\xi_2 = n_XM_X/s = 3.6639\times 10^{-9}\,{\rm GeV}\Omega_Xh ^2$.

\section{Numerical Monte-Carlo Procedure}

The present analysis utilizes a Monte-Carlo approach to the problem.
At each time step of the order of $10^4$ non-thermal nucleons and
$6\times 10^3$ $\gamma$-rays are injected into the thermal plasma.
Here there initial energies are randomly drawn, in the case of nucleons, 
from a kinetic energy distribution function generated by PYTHIA, and in the
case of photons, from the distribution function given in Eq. D.8. Mesons,
which to a large part thermalize before interacting are injected in their
number ratios predicted by PYTHIA. The subsequent evolution of
each nucleon is followed by a Monte-Carlo (random probabilistic)
sequence of events. Energy
losses of nucleons are determined by
a competition between nuclear scattering (spallation) processes
(App. E.3 - E.5), in which a nucleon may loose a good part of its energy,
and continuous energy losses, such as Coulomb stopping, Thomson scattering,
and magnetic moment scattering (App. D.1). The
probability of survival of a nucleon (nuclei) $k$
against nuclear scattering while it looses continuously energy
between energy $E_i$ and $E_f$ is given by
\begin{equation}
P_k(E_i\to E_f) = {\rm exp}\biggl(-\int_{E_f}^{E_i}\frac{{\rm d}E}
{l_N^k(E) |{\rm d}E/{\rm d}x|_c}\biggr)\,
\label{prob}
\end{equation}
where $|{\rm d}E/{\rm d}x|_c$ is the energy-dependent
continuous energy loss per unit path length and $1/l_N^k = 
\sum_i\sigma_{ki\to ...}n_i$ is the inverse of the nucleon (nuclei)
mean free path against nuclear scattering. This probability distribution
is mapped in the analysis by utilizing random numbers. Further, the
evolution of any secondary energetic nucleons due to nuclear scattering
is followed as well. The thermalization of nucleons is followed down to
energies below which inelastic processes are not anymore possible. The
remainder of the thermalization occurs quickly through elastic processes
and does not further influence the BBN yields.
In Fig.~\ref{fig_comp_np} an approximation to the integrand in 
Eq.~\ref{prob} may be seen for protons and neutrons at various cosmic
temperatures.

Photons are treated in a similar way. Each injected photon and the by it
produced secondaries are followed in a Monte-Carlo way until all
photons have dropped below the lowest photodisintegration threshold.
The number of nuclei (nucleons) $A$ per decaying particle $X$
produced due to all $\gamma$-rays created as a consequence of the
decay of $X$ are computed via
\begin{equation}
\frac{N_A}{\rm Decay} = 
\biggl(\sum_i \frac{{\rm d}N(E_{\gamma})}{{\rm d}E_{\gamma}^i}
\Delta E_{\gamma}^i\biggr)
\biggl(\sum_B n_B\sum_j\sigma_{B\gamma\to A...}c\tau_j(E_{\gamma}^j) -
n_A\sum_j\sigma_{A\gamma\to C...}c\tau_j(E_{\gamma}^j)\biggr)\, .
\end{equation}
Here the sum over index $i$ runs over the energy spectrum
of the initial (after CMBR cascade) $\gamma$-ray spectrum, 
$\sigma$ are photodisintegration cross sections,
$c$ is the speed of light, and $\tau_j(E_{\gamma}^j)$ are $\gamma$-ray
survival times against the processes discussed in App. D.2.a, D.2.c, D.2.d,
and D.3.  The sums over $i$ and $j$ are different due to the inclusion
of secondary photons. In particular, a photon with initial energy 
$E_{\gamma}^i$ may
produce secondary photons $j$ due to the processes described in
App. D.2. The sum over $j$ thus includes all $\gamma's$, primary and
secondary, tertiary, ... as long as they are above the photodisintegration
threshold. Individual survival times are generated by a Monte Carlo
and the $\gamma$-ray survival probability 
$P_{\gamma}(E_{\gamma}) = {\rm exp}(\tau c/l_{\gamma}(E_{\gamma}))$,
with $l_{\gamma}(E_{\gamma})$ the $\gamma$ mean free path.
Finally, it is noted that for accuracy reasons the initial first-generation
photon spectrum Eq. D.8 is sampled by an equal amount of photons
above and below the photodisintegration threshold for \he4. This was
necessary as otherwise much larger ($\gg 6\times 10^3$) numbers of
photons would have to be followed in order to obtain accurate result.

\section{Kinematic Relations}

For the convenience of the author this appendix presents some kinematic
relations required to convert cross section data presented in terms
of center-of-mass (CM) quantities to cross section data in the 
laboratory (L) frame of reference. For the energy
gain of particle 2 (initially at rest) in elastic $2\to 2$ scattering,
and given the scattering angle in the CM frame, 
one finds
\begin{equation}
\Delta E_{L,2} = \gamma_{v_{\rm CM}}^2v_{\rm CM}^2M_2(1 - {\rm cos}\,\Theta_{\rm CM})
\end{equation}
where 
$v_{\rm CM} = p_L^1/(M_2 + M_1 + E_{L,{\rm kin}}^1)$,
$E_{\rm CM, 2} = \gamma_{v_{\rm CM}}M_2$, 
and
$|p_{\rm CM}| = v_{\rm CM}\gamma_{v_{\rm CM}}M_2$. 
(Note that $|p_{\rm CM}|$ is
invariant during elastic scattering.)
Concerning
inelastic endothermic
$2\to 2$ scattering reactions $1 + 2\to 3 + 4$ with $2$ initially at rest
(e.g. reactions 77, 82, 85-88) the  
threshold in the laboratory system is given by
$E_L^1 = \Delta m (1+m_1/m_2)$ with
$\Delta m = (m_3+m_4-m_1-m_2)$, the mass differences.
During 
inelastic $2\to 2$ scattering with particle $2$ initially at rest, 
$|p_{\rm CM}|$ is not anymore invariant 
due to conversion of kinetic energy to rest energy.
Utilizing the above given (before scattering) 
$v_{\rm CM}$,$|p_{\rm CM}|$ relations one may solve the implicit
equation $E_{\rm CM}^1 + E_{\rm CM}^2 = E_{\rm CM}^3 + E_{\rm CM}^4$ 
for $p^{\prime}_{\rm CM}$, with
$E_{\rm CM}^3$,and $E_{\rm CM}^4$ containing the after scattering 
modified $p^{\prime}_{\rm CM}$.
Provided one has $\Theta_{\rm CM}$ the CM scattering angle,
the laboratory energy of particles 3 and 4 are determined 
by 
$E_L^3 = \gamma_{v_{\rm CM}}(E_{\rm CM}^3 + v_{\rm CM}|p^{\prime}_{\rm CM}|{\rm cos}\,\Theta_{\rm CM})$
and similarly for particle $4$ with $v_{\rm CM}\to -v_{\rm CM}$. 

\section{Non-Thermal Electromagnetic Interactions}
\label{apx:}

\subsection{Electromagnetic Stopping of Non-thermal Nucleons}

\subsubsection{Coulomb interactions}

Stopping of charged protons and nuclei by Coulomb
scatterings off ambient electrons (and positrons) as well as excitation
of collective plasma modes (involving a larger number of electrons) is
particularly important. In the early Universe single particle scatterings
off electrons and positrons dominate. Since many of these occur with only
small momentum transfer (i.e. small-angle scatterings)
the process may be regarded as a continuous energy loss for 
protons and nuclei. Following Jackson~\cite{Jack}, Gould~\cite{Gould},
and own calculations the energy loss per unit path length traveled
may be found
\begin{equation}
\frac{{\rm d}E_C}{{\rm d}x} = \frac{Z^2\alpha}{v^2}\omega_p^2 
\biggl({\rm ln}\biggl[\frac{0.76v}{\omega_p b}\biggr]
+\frac{1}{2}v^2\biggr)\, ,
\label{Eq:dEcdx}
\end{equation}
where
\begin{equation}
\omega_p^2 = \frac{4\pi n_{e ^{\pm}}\alpha}{m_e}\, ,
\end{equation}
is the square of the plasma frequency, and
\begin{equation}
b = {\rm max}\biggl[\frac{Z\alpha}{\gamma m_e v^2},\frac{1}{\gamma m_e v}
\biggr]\, .
\end{equation}
In these expressions $Z$, $v$, and $\gamma$ refer to electric charge number,
velocity, and Gamma-factor of the nuclei, respectively, $m_e$ is the electron
mass, $\alpha$ the fine structure constant, and $n_{e ^{\pm}}$ the total
number density of electrons and positrons. This latter is given by
\begin{equation}
n_{e ^{\pm}} \approx \bigl(n_{e,\rm net}^2 + n_{e,\rm pair}^2\bigr)^{1/2}\quad
{\rm with}\quad n_{e,\rm pair} = \frac{1}{2}\biggl(\frac{2}{\pi}\biggr)^{3/2}
\bigl(m_eT\bigr)^{3/2}e^{-m_e/T}\biggl(1 + \frac{15T}{8m_e}\biggr)\, ,
\label{Eq:ne}
\end{equation}
where $T$ is cosmic temperature. Here $n_{e,\rm net}$ refers to the
electrons needed for charge neutrality of the Universe. 
Eq.~(\ref{Eq:ne}) is only accurate in
the $T/m_e\ll 1$ limit. Note that the second term in Eq.~(\ref{Eq:dEcdx})
derives from the magnetic moment scattering contribution of the electron 
absent in the Rutherford scattering cross section but present in the correct
Mott scattering cross section. Note also that, in contrast to the findings
of Ref.~\cite{RS} it is found that Eq.~(\ref{Eq:dEcdx}) is also valid in the
extremely relativistic limit up to $\gamma\simle M_p/m_e$, where $M_p$ is
the proton mass. Nevertheless, an important modification to
Eq.~(\ref{Eq:dEcdx}) occurs for small nuclear velocities and large
temperatures when the nuclear velocity $v$ falls below the typical thermal 
electron velocity $v_e$. In this case the Coulomb stopping power 
decreases by a factor $v_e/v^3$~\cite{RS}, a very larger factor $500$ for,
example, for 
tritium nuclei of $10\,$MeV kinetic energy at a temperature 
of $T = 20\,$keV. In order to obtain an accurate result for the stopping
length in this regime an adequate thermal average over the electron velocity
distribution has to be performed, as Coulomb stopping is dominated by 
Coulomb collisions off slow electrons. After a lengthy calculation the author
finds
\bea
{{\rm d}E_C\over {\rm d}x}
={4\pi (Z\alpha )^2\over m_e v} 
n_{e^{\pm}}
\Biggl({\rm erf}(y)\biggl[{\Lambda\over v}
\biggl(1-{T\over 2 m_e}\biggr)
+ {\frac{1}{2}}\biggl(v - \frac{1}{v}\frac{T}{m_e}\biggr)\biggr] 
- \frac{2}{\sqrt{\pi}}e^{-y^2}\frac{y}{v}
\biggl(\Lambda\biggl(1-\frac{T}{2m_e}+\frac{v^2}{8}+
\frac{m_e}{T}\frac{v^4}{8}\biggr)-\frac{T}{2m_e}\biggr]\biggr)
\label{Eq:dEcdx2}
\eea
for the Coulomb stopping which includes a thermal average over a
Maxwell-Boltzmann distribution of electron (positron) velocities. Here
$y = \sqrt{m_ev^2/2T}$ such that for $y\gg 1$ one recovers the
result Eq.~(\ref{Eq:dEcdx}) where the velocities of all electrons are
below that of the nucleus. The term $\Lambda$ in Eq.~(\ref{Eq:dEcdx2})
denoted the logarithm in Eq.~(\ref{Eq:dEcdx}).
Note that in Eq.~(\ref{Eq:dEcdx2}) an
expansion has been performed to lowest non-trivial order in the small
parameters $v$ and $\sqrt{T/m_e}$, where both have been regarded to be
of the same order.

\subsubsection{Thomson scattering}

Energetic protons may loose energy to the CMBR by Thomson scattering,
a process which is particularly efficient at high CMBR temperature.
Due to the large number of CMBR photons many scatterings are involved
and the energy loss of protons due to Thomson drag may also be treated
as a continuous energy loss. Following Reno \& Seckel~\cite{RS} and own
calculations one finds
\begin{equation}
\frac{{\rm d}E_{\rm Th}}{{\rm d}x} = \frac{32\pi}{9}\frac{\pi^2T^4}{15}
\frac{\alpha^2}{M_p^2}\gamma_p^2v_pZ^4\, ,
\label{Eq:dEthdx}
\end{equation}
for the energy loss per unit length where $\alpha$ and $T$ are fine structure
constant and CMBR temperature, $M_p$ the proton mass, $\gamma_p$ and $v_p$ its
gamma factor and velocity, and $Z$ its electric charge. Note that
Eq.~(\ref{Eq:dEthdx}) is applicable for protons of arbitrary 
relativity.

\subsubsection{Magnetic moment scattering}

Neutrons may loose energy due to scattering off electrons and positrons
by virtue of their anomalous magnetic moment. This process is only important
at higher temperature $T\sim 50\,$keV due to the larger numbers of $e^{\pm}$.
By aid of a calculation employing the
relevant interaction amplitude one may find
\begin{equation}
\frac{{\rm d}E_n}{{\rm d}x} = \frac{3\pi\alpha^2\kappa^2m_e}{M_n^2}
\gamma_n^2v_n^2n_{e ^{\pm}}\, ,
\label{Eq:dEndx}
\end{equation}
where $\kappa=-1.91$ is the anomalous magnetic dipole moment and all other
notation is as before. Eq.~(\ref{Eq:dEndx}) applies for non-relativistic as
well as relativistic neutrons.

\subsection{Electromagnetic Cascades Induced by 
$\gamma$-rays and Charged Particles}

The injection of energetic electromagnetically interacting particles in the
early Universe leads to a cascade on the CMBR with $\gamma$-rays 
pair-producing ($\gamma + \gamma_{\rm CMBR}\to e^+ + e^-$) and the produced
electrons (positrons) inverse Compton scattering 
($e^{\pm}+\gamma_{\rm CMBR}\to e^{\pm} + \gamma$) off the CMBR 
photons~\cite{AKV85,ZS89,ZS90,PSB95,KM95}.
This cascade is very rapid due to the large number of CMBR photons.
Only when $\gamma$-ray energies have fallen below that for pair production
$E_{\gamma} < m_e^2/E_{\rm CMBR}$ interactions on protons and nuclei become
important. The spectrum of these ``breakout'' photons below the pair production
threshold depends essentially only on the total electromagnetically 
interacting energy $E_0$ injected above $E_C$. It has been found to
be well approximated by~\cite{PSB95,KM95}
\begin{equation}
\frac{{\rm d}N_{\gamma}}{{\rm d}E_{\gamma}} = \left\{
  \begin{array}{lcl}
\displaystyle
K_0\biggl(\frac{E_{\gamma}}{E_X}\biggr)^{-1.5} &{\rm for}  & E_{\gamma}< E_X
\\*[4mm]
\displaystyle 
K_0\biggl(\frac{E_{\gamma}}{E_X}\biggr)^{-2}&{\rm for}  &  E_C \geq E_{\gamma}\geq E_X 
  \end{array} \right.
\quad ,
\label{Eq:spectrum}
\end{equation}
where
$K_0 = E_0/(E_X^2[2+{\rm ln}(E_C/E_X)])$ is a normalization constant such that
the total energy in $\gamma$-rays below $E_C$ equals the total energy $E_0$
injected. Following the analysis of Ref.~\cite{KM95} 
a value of $E_C\approx m_e^2/22T$
is employed in the present analysis with $E_X\approx 0.03E_C$~\cite{ZS89}.
These values are very close to those ($E_C\approx m_e^2/23.6T$ and 
$E_X\approx 0.0264E_C$) advocated by Ref.~\cite{PSB95}. 
Subsequent interactions of these ``break-out'' photons include photon-photon
scattering ($\gamma +\gamma_{\rm CMBR}\to \gamma + \gamma$) mainly
redistributing the energy of energetic $\gamma$-rays right below energy $E_C$,
Bethe-Heitler pair production 
($\gamma + p({}^4{\rm He})\to p({}^4{\rm He}) + e^- + e^+$),
Compton scattering ($\gamma + e^-\to \gamma + e^-$) off thermal electrons,
with the produced energetic $e^-$ inverse Compton scattering to generate
further low-energy $\gamma$-rays, as well as nuclear photodisintegration.
All these processes are included in the analysis and their detailed
treatment is described below.

\subsubsection{Bethe-Heitler pair production}

The cross section for $e^{\pm}$ pair production by $\gamma$-rays
of energy $E_{\gamma}$
scattering off protons and helium nuclei at rest is given by
\begin{equation}
  \sigma_{\rm BH}\approx \frac{3}{8}\frac{\alpha}{\pi}\sigma_{\rm Th}\biggl(
\frac{28}{9}{\rm ln}
\biggl[\frac{2E_{\gamma}}{m_e}\biggr] -\frac{218}{27}\biggr)Z^2\, , 
\label{sigBH}
\end{equation}
where $\sigma_{\rm Th}$, $\alpha$, $m_e$, and $Z$ are Thomson cross section,
fine structure constant, electron mass, and 
nuclear electric charge, respectively.
The expression is valid in the regime 
$1\ll E_{\gamma}/m_e\ll \alpha^{-1}Z^{-1/3}$. For energies in the range
$2m_e < E_{\gamma} < 4\,{\rm MeV}$ the cross section is roughly approximated
as constant at the value of Eq.(\ref{sigBH}) at $E_{\gamma} =4\,$MeV.  
The energy of the produced $e^{-}$ ($e^{+}$) is approximated
by a flat probability distribution within the range $[m_e,E_{\gamma}-2m_e]$
with the antiparticle $e^{+}$ ($e^{-}$) carrying the remainder of 
$E_{\gamma}$.

\subsubsection{Inverse Compton scattering}

Electrons and protons produced during the dominant Bethe-Heitler process
may upscatter CMBR photons to $\gamma$-ray energies by inverse Compton 
scattering. As these $\gamma$-rays may later photodistintegrate
nuclei, a fairly accurate treatment of inverse Compton scattering 
is required. Here the exact inverse Compton scattering rate is unimportant
since it is always large compared to the Hubble expansion.
For the cumulative probability distribution that an $e^{-}$ ($e^{+}$)
of energy $E_e =\gamma_em_e$ upscatters a CMBR photon to 
energy $E_{\gamma} = 4\gamma_e^2Tx$, where $T$ is CMBR temperature, one may
find
\begin{equation}
P_{ic} \approx e^{-x}\biggl(1+\frac{1}{4}x-\frac{1}{4}x^2\biggr)
+\Gamma(0,x)\biggl(\frac{1}{4}x^3 - \frac{3}{4}x\biggr)\, ,
\label{Pic}
\end{equation}
with
\begin{equation}
\Gamma (0,x) = \int_x^{\infty}dy\, e^{-y}/y\,
\end{equation}
is an incomplete Gamma function.
Note that for the evaluation of Eq.~(\ref{Pic}) the approximation that
the $e^{\pm}$ are ultrarelativistic, i.e. $\gamma_e\gg 1$, and that
the scattering process occurs in the Thomson regime, 
i.e. $12\gamma_eT/m_e\ll 1$, have been made. These approximations are
appropriate since only ultrarelativistic $e^{\pm}$ may produce $\gamma$-rays
by inverse Compton scattering on the CMBR which are energetic enough to
photodisintegrate nuclei, and since $e^{\pm}$ produced during Bethe-Heitler
pair production are not energetic enough for the inverse Compton scattering
to proceed in the Klein-Nishina limit.
Furthermore for the derivation of Eq.~(\ref{Pic}) the Einstein-Bose
occupation number $(e^{-E_{\gamma}}-1)^{-1}$ has been approximated by
$\xi(3)e^{-E_{\gamma}}$ with $\xi(3)\approx 1.2021$ such that the
total number of CMBR photons remains the same. The approximation was
required to find a closed form for $P_{ic}$. By comparison to an accurate
numerical evaluation one finds that the approximation is typically good to
within 3\% - 10\%. Lastly 
one finds for the average scattered CMBR photon energy
a value of $\langle E_{\gamma}\rangle =\gamma_e^2\langle E_{\rm CMBR}\rangle$
with $\langle E_{\rm CMBR}\rangle\approx 2.701T$ the average CMBR photon 
energy. 

\subsubsection{Compton scattering}

The cross section for $\gamma$ rays of energy $E_{\gamma}$
scattering off electrons at rest
in the Klein-Nishina limit $E_{\gamma}\gg m_e$ is given by~\cite{IZ}
\begin{equation}
\sigma_{\gamma e} \approx \frac{3}{8}\frac{\sigma_{\rm Th}}{\omega}
\biggl[{\rm ln}2\omega + \frac{1}{2}+{\rm O}
\Bigl(\frac{\rm ln \omega}{\omega}\Bigr)
\biggr]
\quad \omega\gg 1\, ,
\end{equation}
where $\omega \equiv E_{\gamma}/m_e$ and $\sigma_{\rm Th}$ is the Thomson
cross section. The cumulative probability distribution that the $\gamma$
scatters to energy $E^{\prime}_{\gamma} \leq E^{c}_{\gamma}$ 
during the Compton scattering process
may be derived by utilizing the
results in Ref.~(\cite{IZ}) and is found as
\begin{equation}
P_c(E_{\gamma},E^{\prime}_{\gamma}\leq E^{c}_{\gamma})\approx 
\frac{\frac{1}{2} x^2 - {\rm ln} \frac{1}{x} + {\rm ln}2\omega}{\frac{1}{2} + 
{\rm ln}2\omega} +{\rm O}(1/\omega)\, ,
\end{equation}
where $x\equiv E^{c}_{\gamma}/E_{\gamma}$ and the kinematic limits of $x$
fall in the range $1$ and $1/(1+2\omega)\approx 1/2\omega$. 
The scattered electrons may produce further $\gamma$-rays by inverse
Compton scattering on the CMBR, an effect which is taken into account
in the calculations. 

\subsubsection{$\gamma$-$\gamma$ scattering}

Following the analysis of Svensson \& Zdziarski~\cite{ZS89,ZS90} the total
rate for scattering of a $\gamma$-ray of energy $E_{\gamma}$ on a
blackbody of temperature $T$ is given by
\begin{equation}
R_{\gamma\gamma} = \frac{2^4139\pi^3}{3^65^4}\alpha^4m_e
\biggl(\frac{T}{m_e}\biggr)^6\biggl(\frac{E_{\gamma}}{m_e}\biggr)^3\, .
\end{equation} 
Employing Eq.(3.9) of Ref.~\cite{ZS90} one may compute the cumulative
probability distribution for the energy of the scattered photon 
$E^{\prime}_{\gamma}$ to be below $E^c_{\gamma}$ as
\begin{equation}
P_c(E_{\gamma},E^{\prime}_{\gamma}\leq E^c_{\gamma}) =
\frac{10}{7}x\biggl[1-x+x^2-\frac{1}{2}x^3+\frac{1}{5}x^4\biggr]\, ,
\end{equation}
where $x\equiv E^c_{\gamma}/E_{\gamma}\leq 1$. The energy of the scattered
blackbody photon is given approximately by 
$E^{\prime}_{\gamma,BB}\approx E_{\gamma} - E^{\prime}_{\gamma}$ in the limit
$T\ll E_{\gamma}$ and its effects are treated in the calculations as well.

\subsection{Photodisintegration}

Photodisintegration of the light-elements by $\gamma$-rays generated
as a consequence of relic particle decays plays a particularly important
role in the determination of BBN yields, 
as noted early on~\cite{Lindley}. It is
only effective at temperatures below $T\simle 6\,$keV (cf. Fig.~\ref{destEM}), 
since at higher
temperatures $\gamma$-rays of sufficient energy $E_g\approx 2\,$MeV to
photodisintegrate \h2, \li7, and \be7 predominantly pair produce on the
abundant CMBR photons (cf. App. D.2). Below this approximate 
cosmic temperature pair-production is kinematically forbidden such that
photodisintgeration becomes probable. The competing Bethe-Heitler
pair production, Compton scattering and $\gamma\gamma$ scattering
were described in the previous subsections.

The (lowest) photodisintegration thresholds for the various light nuclei
synthesized during BBN
\h2,\h3,\he3,\he4,\li6,\li7, and \be7 are given by
$2.225\,$MeV, $6.257\,$MeV, $5.493\,$MeV, $19.814\,$MeV, $3.699\,$MeV,
$2.467\,$MeV, and $1.587\,$MeV, respectively. The present analysis
utilizes photodisintegration data as parametrized by Ref.~\cite{Cyburt}
with a few of these parametrizations changed to produce improved fits
to the available reaction data. For original references the reader is
referred to the references in Ref.~\cite{Cyburt}.

In addition, the reactions 23-26, and 30-33, not considered in 
Ref.~\cite{Cyburt}, have been included in the present analysis. Here the
following parametrizations of reaction rate data~\cite{Varl,7ligp,Koti}
have been adopted:
\vskip 0.1in
\noindent
Reaction 23~\cite{Varl}:
\begin{equation}
\sigma (E_{\gamma}) = 3.8 {\,\rm mb\,}
Q_0^{2.3} (E_{\gamma}-Q_0)/E_{\gamma}^{3.3} 
+ \Bigl(2.1{\,\rm mb\,} 
Q_1^{1.5}(E_{\gamma}-Q_1)/E_{\gamma}^{2.5}\Bigr)\Theta (E-Q_1)
\end{equation}
with
$Q_0 = 8.725\,$MeV and $Q_1 = 23\,$MeV and $\Theta (x)$ the step function,
\vskip 0.1in
\noindent
Reaction 24~\cite{7ligp,Varl}:
\begin{equation}
\sigma (E_{\gamma}) = 10.8 {\,\rm mb\,} Q^2 (E_{\gamma}-Q)^{1.2}/E_{\gamma}^{3.2}
\end{equation}
with
$Q = 9.98\,$MeV, and
\vskip 0.1in
\noindent
Reaction 25~\cite{Koti}:
\begin{equation}
\sigma (E_{\gamma}) = 1.44\times 10^3 {\,\rm mb\,} Q^{20}
(E_{\gamma}-Q)^{2.4}/E_{\gamma}^{22.4}
\end{equation}
with $Q = 22.28\,$MeV.
Reaction 26 has been parametrized
as reaction 25 but with
$Q = 23.05\,$MeV since no experimental data exists.
Similarly, since no reaction rate data exists for
the cross sections of reactions 30-33 they have been approximated by the data 
of their respective mirror reactions 23-26 but with binding energies 
appropriately replaced by, $Q = 7.08\,$MeV, $Q = 10.68\,$MeV, $Q = 22.17\,$MeV,
and $Q = 21.40\,$MeV for reactions 30,31,32, and 33, respectively.

\section{Non-Thermal Hadronic interactions}
\label{apx:1}

\subsection{Mesonic Induced Charge-exchange reactions}

Injection of mesons in the form of metastable pions and kaons may induce the
charge-exchange reactions 34-43 to convert protons to neutrons and 
vice versa~\cite{RS}. 
As essentially each neutron is incorporated into a helium nuclei at
$T\approx 80\,$keV, mesons may thus effect the \he4 abundance. Since more
protons are converted to neutrons than vice versa, injection of mesons leads
to an increase in the \he4 abundance. Mesonic charge-exchange reactions
are particularly important at fairly high temperatures. At $T\approx 1\,$MeV,
for example, a fraction $\sim 0.01$ of all mesons induce charge exchange with
the remainder decaying. In contrast, this fraction drops to $\sim 10^{-5}$
at $T\approx 100\,$keV. The present analysis follows the treatment of 
Ref.~\cite{RS} and the reader is referred to this study for details.

The mesons may be due to a variety of processes.
It is well known that the hadronic decay (or annihilation) 
results in a multitude of mesons. However, under certain circumstances
even electromagnetic- and weak- decays may yield mesons. This may be 
due to decays involving $\tau^{\pm}$ in the final state, which have a 
branching ratio to pions. Similarly injection of high-energy 
$E_{\nu}\sim 100\,$GeV neutrinos at $T\sim 1\,$MeV may pair-produce
$\pi^{\pm}$ on the neutrino background. Pair-production of pions by 
$\gamma$-scattering on the CMBR~\cite{KKM} is unlikely as the Compton
cross section is typically a factor $\sim 300$ larger.

Production of mesons by electromagnetically or weakly interacting
particles has not been included in the present study. In the case
of hadronic decays their effect is subdominant, and in the
case of electromagnetic- and weak- decays the analysis is model-dependent
on the decay products.

\subsection{Nucleon-Antinucleon Annihilations}

Hadronic decays lead to the production of nucleons and anti-nucleons.
It is assumed here that the total baryon number is conserved during
the decay.
At higher temperatures ($T\simge 90\,$keV for $n$ and $\bar{n}$ and
$T\simge 20\,$keV for $p$ and $\bar{p}$) they thermalize by electromagnetic
energy losses (cf. App. D.1) with antinucleons annihilating at rest
thereafter. The combination of injection of nucleons (with a predefined
$n/p$-ratio, typically close to one) and annihilation of antinucleons on
the pre-existing nucleons may lead to an increase of the
neutron-to-proton ratio and the \he4 abundance resulting from BBN~\cite{RS}.
Antinucleons mostly annihilate on protons due to the small thermal
$n/p$-ratio and due to Coulomb enhancement of the $\bar{p}p$ annihilation
cross section. The annihilated protons are replaced by the 
injected nucleons which typically come in a ratio $n/p\approx 1$ thus
yielding an effective increase of the $n/p$-ratio. 
Given a typical injected meson-to-(anti-)baryon ratio of $\sim 20$ 
in hadronic decays
the effects of nucleon anti-nucleon injection are far more important 
than those of mesonic charge exchange reactions. This holds particularly 
true at lower temperatures $T\simle 300\,$keV. The present study uses
the reaction rate data as given in Ref.~\cite{RS}.

When electromagnetic stopping becomes less dominant at lower temperatures
thermalization of energetic nucleons occurs partially due to 
nucleon-nucleon scattering and nuclear 
spallation (cf. App. E.3,E.4,E.5). Antinucleons, on the
other hand, never thermalize but rather annihilate during one of their 
first interactions with thermal protons or helium-nuclei. As an energetic
nucleon has of the order of $\sim 20$ $N-N$ scattering events before
dropping below the \he4 spallation threshold it's probability to inflict
a \he4 spallation event is factor $\sim 10$ larger than that of an
antinucleon. It has been therefore approximated that antinucleons
do never spall \he4, leading to an about 10\% underestimate in \h2, \h3, and
\he3 production due to particle decay.

\subsection{Elastic $N$-$N$ scattering}

Elastic nucleon-nucleon scattering and nucleon-\he4 scattering of
energetic neutrons and protons off the thermal protons and \he4 is
important in the nucleon thermalization process in particular at lower
temperature ($T\simle 90\,$keV for $n$'s and
$T\simle 20\,$keV for $p$'s) and when the nucleons are not too energetic,
i.e. $E\simle 1\,$GeV. This may be seen in Fig.~\ref{fig_comp_np}.
In contrast, very energetic nucleons, $E\gg 1\,$GeV,
are stopped at lower temperatures mostly by inelastic scatterings of
protons and \he4 (cf. App. E.4, and E.5) and at higher temperatures either
by Thomson scattering off CMBR photons in the case of protons (cf. App. D.1.b)
or magnetic moment scatterings off thermal $e^{\pm}$ in the case of neutrons
(cf. App. D.1.c). Elastic scatterings off protons also produce secondary
energetic protons which are taken into account in the present study.

For the elastic $p$-$p$, $n$-$p$, $p$-\he4, and $n$-\he4 scattering cross
sections we take the data as compiled by Ref.~\cite{Meyer}. This
compilation compares well with that given in the more recent compilation
given in Ref.~\cite{particle}. It is noted here that due to approximate
isospin invariance the $p$-\he4 and $n$-\he4 cross sections are almost
equal. To calculate the energy loss in $p$-$p$ and $n$-$p$ scattering
events, an against nuclear data tested algorithm presented
in Ref.~\cite{SAID} is
used. Each scattering event is treated by a Monte Carlo. This algorithm
predicts an almost uniform probability distribution for fractional
energy loss $\Delta E_k/E_k$ for $\simle 250\,$MeV protons progressing 
smoothly towards a bimodal probability distribution (either forward scattering,
i.e. $\Delta E_k/E_k\approx 0$ or backward scattering, 
i.e. $\Delta E_k/E_k\approx 1$)
when the $p$ kinetic energy is increased towards $2\,$GeV. Neutron-proton
scattering follows a similar though less well-defined trend. For kinetic
energies $E_k > 2\,$GeV the probability distributions of $2\,$GeV nucleons
are utilized.

Nucleon-\he4 scattering follows a very different pattern with \he4 recoil
energies to high probability very small. Given data by Ref.~\cite{he4recoil}
the typical \he4 recoil energy is approximated at $5\,$MeV in each event.
It has been verified that a change in this recoil energy has negligible
effect on the BBN yields. For a more detailed discussion
on the \he4 recoil energies which
play a role in \he4-\he4 fusion reactions 87 and 88  
the reader is referred to
App. E.6.  

\subsection{Inelastic $N$-$N$ scattering}

For nucleon-nucleon scattering with nucleon kinetic energies in the GeV range
scattering processes are mostly inelastic accompanied by the production of pions.
Treating these processes properly is important not only for a determination of
the typical nucleon energy loss, but also because scattering processes may easily
convert protons to neutrons and vice versa. In fact, due to a large asymmetry in the
cross sections the conversion of energetic protons to neutrons occurs more often
than the inverse of this process. This has implication for spallation processes,
increasing \h2,\h3,\he3, and \li6  yields, as neutrons
are stopped 
by scattering off protons and \he4 for temperatures $T\simle 90\,$keV,
whereas $E_p\simle 1\,$GeV protons mostly loose their energy via Coulomb stopping. 
This study uses the inelastic $N$-$N$ cross sections reactions 52 - 73 
in Table 1
as compiled by Ref.~\cite{Bystricky},
where some of the proposed fitting functions had to be modified by the author the
correctly account for the data. Concerning the total average energy loss $f_{\pi}$
into pion mass and kinetic energy in inelastic scatterings experimental data
by Ref.~\cite{Thomas} at $E_p = 790\,$MeV and a simulation with 
Pythia at $E_p = 53\,$GeV 
has been fitted logarithmically to obtain $f_{\pi}\approx 0.2 + 0.08{\rm ln}(E/0.8{\rm GeV})$ 
The remainder of the energy has been split in proportions (0.15,0.85) between the
outgoing nucleons roughly equal to the average energy split for energetic elastic
collisions. 
For the nucleon energy loss in the inelastic spallation processes 
discussed in App. E.5 it has been assumed that the outgoing nucleons are missing
the binding energy necessary for the spallation process, and the remaining energy is shared
between primary
and secondary nucleon in proportion (0.75,0.25). This corresponds to the average values for elastic
$N$-$N$ scattering in the several hundred of MeV range. 

\subsection{Nuclear spallation}

\bef
\showone{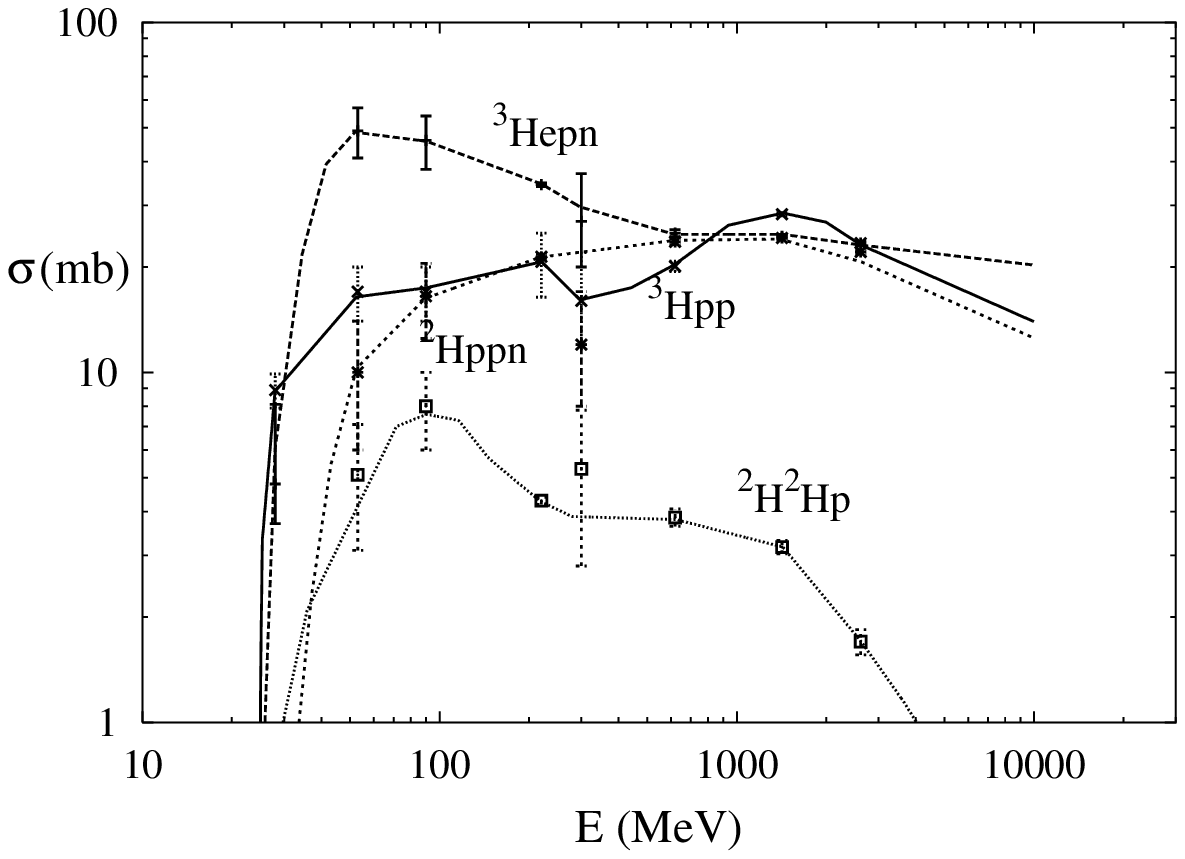}
\caption{Compilation of the available \he4 spallation data as a function
of nucleon kinetic energy (with \he4 at rest) for the
spallation reactions 74,75,76, and 78. The lines joining independent data
points (74 - solid; 75 - long-dashed, 76 - short-dashed, 78 - dotted) 
represent the fit used in the present analysis. Refer to the text
for references.}
\label{fig_cross}
\eef

Aside from \he4 photodisintegration which is operative only at lower
cosmic temperatures $T\simle 3\,$keV \he4 spallation reactions 
at temperatures $T\simle 90\,$keV induced
by energetic nucleons are the main source of additional \h2, \h3, \he3,
as well as \li6 (through nonthermal fusion reactions induced by energetic
3-nuclei, cf. App. E.6) in BBN with decaying particles. It is therefore 
important to employ accurate reaction cross section data. The 
\he4-spallation reactions 
74 - 83 in Table 1 have been included in the analysis. 
Experimental data on these reactions has been obtained at nucleon energies
$E = 28\,$MeV~\cite{Wicker}, $53\,$MeV~\cite{Cairns}, 
$90\,$MeV~\cite{Tannenwald}, 
$220\,$MeV~\cite{Abdullin}, $300\,$MeV~\cite{Innes}, 
$620\,$MeV~\cite{Blinov2}, 
$1.42\,$GeV~\cite{Glagolev}, and
$2.61\,$GeV~\cite{Glagolev}, respectively, 
and has been interpolated in between. This data and the interpolation is shown
in Fig.~\ref{fig_cross}.
The data also includes the possible
production of additional pions in the spallation process. Note that
\he4-spallation data induced by either neutrons or protons has been
included on equal basis in the compilation. Here reactions induced
by neutrons, such as $\He4(n,np)\H3$, have been associated with their
corresponding mirror reactions induced by protons, i.e.
$\He4(p,np)\He3$. This is theoretically and experimentally justified due
to approximate isospin invariance, as long as Coulomb effects are
negligible, in particular, for nucleon energies larger than a few MeV.
For the energies of nucleons and nuclei's after the spallation the reader
is referred to App. E.4 and E.6. The spallation of energetic freshly 
synthesized \li6 reaction 84 has also been included~\cite{Elwyn} 
but plays only a minor role for the results.

\subsection{Non-Thermal Fusion Reactions}

Non-thermal fusion reactions are particularly important for the
formation of \li6 which during thermal BBN is produced only in
very small amounts. When relic particles decay during or after
BBN \li6 is therefore the first element
which becomes significantly perturbed compared to its SBBN abundance.
This has been shown first in the case of hadronic decays in 
Ref.~\cite{Dimo} and later in the case of electromagnetic decays in 
Ref.~\cite{J00}.
The most efficient reaction sequence is initiated
by either the spallation $\He4(N, ...$
or photodisintegration $\He4(\gamma, ...$ of \he4 to produce energetic
\h3 and \he3 nuclei. These later may then fuse on ambient \he4-nuclei
to form \li6 via $\H3(\He4 ,n)\Li6$ and $\He3(\He4 , p)\Li6$. If only a
small fraction $\sim 10^{-4}$ (cf. Fig.~\ref{fig_6Li})
of these energetic 3-nuclei fuse an
observationally important \li6 abundance may result. Energetic 3-nuclei
in the tens of MeV range are essentially all stopped by 
Coulomb interactions. Since this process is less efficient at higher
energy (i.e. $(1/E_3)dE_3/dx\propto E_3^{-2}$; cf. App. D.1.a) and since
the fusion cross section does not drastically fall with energy the formation
of \li6 is very sensitive towards the high-energy tail of 
produced 3-nuclei. It will be seen that the lack of knowledge of this tail
actually introduces some uncertainty in the \li6 yield.

In this paper the $\H3(\He4 ,n)\Li6$ and $\He3(\He4 , p)\Li6$
cross sections as determined with the
aid of reverse reaction rate data by Ref.~\cite{Sihvola} 
and Ref.~\cite{Cyburt}, respectively, have been adopted. Here the
first reaction has a threshold in the \he4 rest frame of
$E_{th} = 8.39\,$MeV,
whereas the second reaction has $E_{th} = 7.05\,$MeV. Cross sections
for the spallation and photodisintegration reactions are taken as described
in App. E.5 and App. D.3, respectively. Concerning the ``recoil'' energy
distributions of the 3-nuclei in the case of photodisintegration it is
given by $E_{3} = (E_{\gamma}-E_{\gamma ,th}^{\He4})m_N/(m_N + m_{3})$
for \h3 nuclei, 
whereas for spallation one has to resort to experimental data.
Data has been published for the $\He4(n,np)\H3$ reaction~\cite{Tannenwald}
at $E_n = 90\,$MeV in the \he4 rest frame
and more recently for the $\He4(p,pp)\H3$ and $\He4(p,np)\He3$ 
reactions~\cite{Blinov} at $E_p = 220\,$MeV. Data at higher energies
exists as well~\cite{3distr} which lies
above the threshold of pion production during
the spallation process. In general, the energy distributions of 3-nuclei
seem almost independent of incident nucleon energy though care has to be 
taken in the comparison
since sometimes only partial (quasi-free-scattering) distributions
are shown (as in the case of Ref.~\cite{3distr}).
In this study the complete distributions as given in Ref.~\cite{Blinov}
are utilized (in binned form) for both the $\He4(p,...$ reactions and
their $\He4(n,...$ corresponding mirror reactions. 
The utilization of reaction rate
data of the corresponding mirror reactions seems not only theoretically
justified~\cite{Meyer,Chad} but also when experimental data of 
Ref.~\cite{Tannenwald} and Ref.~\cite{Blinov} are directly compared.

Some uncertainty in the \li6 yield results from an inaccurate determination
of the few percent high-energy tail in the 3-nuclei energy
distribution function. The highest energy events are particularly difficult
to detect in bubble chamber experiments~\cite{Chad}, such that
the distribution of Ref.~\cite{Blinov} only extends towards 
$E_3\approx 90\,$MeV. In Ref.~\cite{Tannenwald} a fraction of about $2\%$
of all spallation reactions seem to result in \h3 nuclei at the 
kinematically highest allowed energy. This corresponds to backward
scattering and such peaks in the distribution functions are also observed
in \he4-\he4 elastic scatterings~\cite{he4back}. If the four-event backward
scattering peak as observed in Ref.~\cite{Tannenwald} is real, rather than
a fluctuation or measurement error, resulting \li6 yields may increase by
some tens of per cents due to these high-energy 3-nuclei 
(cf. Fig.~\ref{fig_6Li}). However, on 
conservative grounds this study neglects \li6
production due to the very highest energy 3-nuclei.

Energetic 3-nuclei may also be produced by $\He4(n,\H2 )\H3$ and its mirror
reaction. For the energy distributions of this subdominant 3-nuclei channel
reaction data of the compilation/determination in
Refs.~\cite{Meyer,Votta} at proton kinetic energies of $55\,$MeV, $85\,$MeV,
and $156\,$MeV, respectively, have been interpolated. Fusion reactions such
as $\H2(\alpha ,\gamma )\Li6$, $\H3(\alpha ,\gamma )\Li7$, and
$\He3(\alpha ,\gamma )\Be7$ have not been considered as cross sections
involving the emission of a photon are typically only in the $nb$ to
$\mu b$ range. The $N(N,\pi's)\H2$ reaction, on the other hand is
treated as discussed in App.E.4.

Note that in a narrow temperature interval around $T\sim 3- 20\,$keV
Coulomb stopping of not too energetic $E_3\sim 10-30\,$MeV 3-nuclei looses
its efficiency (cf . App. D.1.a) due to the nuclei velocity falling below
a typical electron thermal velocity. In this regime of efficient \li6
synthesis other processes reactions 89 - 91
which contribute to the thermalization of
3-nuclei may become of some limited
importance. They have been included in the calculation by
taking data from Ref.~\cite{Haesner}. At very low temperatures 
$T\simle 1\, eV$ 
energetic \h3 stopping times may become longer
than the \h3 decay time. It is therefore possible that \h3 decays to \he3
before it falls below the \li6 fusion threshold and this effect has been taken
into account.

Fusion reaction between energetic \he4 (produced by $N$-\he4 elastic
scatterings) and thermal \he4-nuclei may, in principle, also
lead to the synthesis of \li6, as well as \li7. 
In practice, however, this process
is subdominant and has therefore not been included. The threshold energies
for $\He4(\alpha ,p)\Li7$ and $\He4(\alpha ,pn)\Li6$ are $37.8\,$MeV
and $49.2\,$MeV, respectively. These are much larger than those for
the fusion of 3-nuclei on \he4. 
Given the data in Ref.~\cite{he4recoil}
the distribution function for recoil energies
of \he4-nuclei in $N$-\he4 elastic scattering may be approximated by an
exponential distribution
$dP/dE_4\approx a\, {\rm exp}(-bE_4)$ with $a\approx 225\,$mb/GeV and
$b\approx 65\,$GeV$^{-1}$, independent of nucleon kinetic energy.
The total cross section for producing $E_4\simge 50\,$MeV recoil \he4
in $N$-\he4 scatterings is thus in the $\sim 0.1\,$mb range. Note that
$E_4\simge 50\,$MeV has been chosen as only for such energies the 
\he4-\he4 fusion cross
section becomes appreciable $\sim 100\,$mb~\cite{alal}.
The cross section of $\sim 0.1\,$mb for the production of \he4 nuclei
energetic enough to produce \li6 by fusion has to be compared to 
the typical total cross sections for $\He4(N,...)\H3$, etc. of $\sim 30\,$mb
times the fraction $\sim 0.3$ of 3-nuclei energetic enough to fuse to \li6.
This yields $\sim 10\,$mb compared to $0.1\,$mb. Nevertheless, \li6 fusion
via \he4-nuclei is enhanced due to a less efficient Coulomb stopping at
$50\,$MeV versus the canonical $10\,$MeV for 3-nuclei, 
resulting in an enhancement factor of
$(50/10)^2/4\approx 6$. Here the $4$ in the denominator comes from the
factor two higher charge of \he4 nuclei compared to \h3 nuclei.
Altogether it is found that energetic \he4 contributes
only of the order $\sim 6\%$ to the final \li6 abundance
This is also not changed by the existence of a sharp backward scattering
peak in $N$-\he4 scattering~\cite{he4back} which, similarly to the
possible existence of a backward peak in the 3-nuclei distribution
(see above) may increase the \li6 synthesis due to energetic \he4 by a factor
$\sim 1.2$.

\section{Thermal nuclear reactions}

Thermal nuclear reactions are treated by the Kawano code updated by the
NACRE~\cite{NACRE} reaction compilation. For the neutron life time
the value $\tau_n = 885.7\,$sec is used as quoted as the 'world' average
in the Particle Data Booklet.
For the fractional contribution
of baryons to the critical density the central
value as determined by three years
of WMAP observations~\cite{WMAP} $\Omega_bh^2= 0.02233^{+0.072}_{-0.091}$
is applied. Given the conversion
$\Omega_b h^2 = 3.65\times 10^7\eta$ this corresponds
to a baryon-to-photon ratio of $\eta = n_b/n_{\gamma} = 6.12\times 10^{-10}$.
With these input values the abundances within a SBBN scenario are found at
\h2/H $= 2.68\times 10^{-5}$, \he3/\h2 $= 0.38$ (\he3/H $= 1.04\times 10^{-5}$)
$Y_p = 0.248$, and \li7/H $= 4.34\times 10^{-10}$, respectively. The \li6
abundance in a SBBN scenario is negligible compared to the observationally
inferred value of \li6 in low-metallicity stars.




\begin{thebibliography}{99}
\bibitem{Cybphoto} Cross section approximated by the 
parametrisation given in Ref.~(\cite{Cyburt}) after comparison
with published data in the 
'Experimental Nuclear Reaction Data Base (EXFOR)'. For the original
references the reader is referred to the EXFOR database
and Ref.~(\cite{Cyburt}).
\bibitem{remark99} The parametrisation as given in Ref.~(\cite{Cyburt})
is adopted. Nevertheless, when reverse reaction rate data from
NACRE~\cite{NACRE} is used to evaluate the cross section 
the parametrisation leads to 
a factor $2-3$ overestimate of the cross section. In contrast, when direct
reaction data assembled by Ref.~\cite{Varl} is used the parametrisation
leads to a factor $2-3$ underestimate of the cross section. Other 
direct reaction rate data, such as of Ref.~\cite{Griff} also does not lead
to a favorable comparison. It is concluded that the adopted
$\Li7(\gamma ,\H3 )\He4$ cross section has large uncertainties. Since
constraints from \li7 are rather weak this has only little impact on
the presented constraints.
\bibitem{remark100} The parametrisation of Ref.~(\cite{Cyburt}) is adopted
with nevertheless negative values of $\sigma$
which occur at larger $E_{\gamma}$ in this parametrisation set to zero. 
Direct reaction data for this reaction does not exist.
As in the case of the reaction $\Li7(\gamma ,\H3 )\He4$ an evaluation of
the cross section by the author utilising
revere reaction rata data from Ref.~\cite{NACRE}
indicates a possible overestimate of the cross section as parametrised in
Ref.~(\cite{Cyburt}). Nevertheless, as constraints from other light elements
are more stringent, this possible overestimate has only negligible influence
on the conclusions.
\bibitem{remark200} An own fit to the data has been used since the fit 
as proposed by Ref.~\cite{Bystricky} is not adequate.
\end{thebibliography}
\end{document}